\documentclass[10pt,leqno]{article}

\usepackage{amsmath}
\usepackage{amssymb}
\usepackage[noend]{algpseudocode}
\usepackage[ruled]{algorithm}
\usepackage{pstricks}
\usepackage{pst-node}
\usepackage{multirow}
\usepackage{amsthm}
\usepackage{float}

\newtheorem{theorem}{Theorem}[section]
\newtheorem{corollary}[theorem]{Corollary}
\newtheorem{proposition}[theorem]{Proposition}

\newtheorem{fact}[theorem]{Fact}
\newtheorem{claim}[theorem]{Claim}
\theoremstyle{definition}
\newtheorem{definition}[theorem]{Definition}
\newtheorem{example}[theorem]{Example}

\newcommand{\conflicts}{\leftrightsquigarrow}
\newcommand{\nll}{\not \ll}
\DeclareMathAlphabet{\mathpzc}{OT1}{pzc}{m}{it}
\newcommand{\g}{\ensuremath{\mathpzc{g}}}
\renewcommand{\l}{\ensuremath{\mathpzc{l}}}
\DeclareMathOperator{\Rep}{Rep}
\DeclareMathOperator{\MIS}{MIS}
\DeclareMathOperator{\LRep}{LRep}
\DeclareMathOperator{\GRep}{GRep}

\author{
  Slawomir Staworko \quad Jan Chomicki\\
  University at Buffalo \\
  \{staworko,chomicki\}@cse.buffalo.edu\\
}
\date{}
\title{
  Priority-Based Conflict Resolution in \\
  Inconsistent Relational Databases\thanks{Research supported by NSF
  Grants IIS-0119186 and IIS-0307434.}
  \footnotetext[2]{UB CSE Technical Report 2005-11} 
}
\begin{document}
\thispagestyle{empty}
\maketitle
\begin{abstract}
We study here the impact of priorities on conflict resolution in
inconsistent relational databases. We extend the framework of
\cite{ArBeCh99}, which is based on the notions of repair and
consistent query answer. We propose a set of postulates that an
extended framework should satisfy and consider two instantiations of
the framework: (locally preferred) \l-repairs and (globally preferred)
\g-repairs. We study the relationships between them and the impact each
notion of repair has on the computational complexity of repair
checking and consistent query answers. 
\end{abstract}
\vspace{20pt}
\noindent

\section{Introduction}
The main purpose of integrity constraints is to express semantic
properties of the data stored in the database. Usually, it is the
database management system that is responsible for maintaining
the integrity of the database. However, in many recent applications
the integrity enforcement becomes a problematic issue. For example in
the data integration setting, even when the data contained by a data
source satisfies the integrity constrains, a different data source
may contribute conflicting information. At the same time data sources
may be autonomous and it may be impossible to modify their contents in
order to remove the conflicts. Integrity constraints may also fail to
be enforced because of efficiency considerations. Finally, in the case
of long running operations, integrity violations may be only temporary
and will be eliminated by further operations.

Typically, the user formulates a query with the assumption that
the database is consistent (i.e. satisfies the integrity constraints).  
A simple evaluation of the query over an inconsistent database may
return incorrect answers. To address this problem Arenas, Bertossi,
and Chomicki \cite{ArBeCh99} proposed the framework of 
{\em consistent query answers}. They introduced the notion of a 
{\em repair}: a consistent database that is minimally different from
the original one. A {\em consistent answer} to a query is an answer
{\em true} in {\em every} repair. 
The framework of \cite{ArBeCh99} is used as a foundation for 
most of the work in the area of querying inconsistent databases
\cite{ArBeCh03,ABCHRS03,CaLeRo03,BrBe03,EFGL03,FuMi05,FuFaMi05,BFFR05}.
\begin{example}\label{ex:cqa}
  Consider a database consisting of two tables $Emp$ and $Mgr$ whose 
  instance $I_0$ can be found in Table~\ref{tab:tab4}.
  \begin{table}[ht]
    \begin{center}
      \begin{tabular}{|c|c|}
        \multicolumn{2}{c}{$Emp$}\\
        \hline
        Name & Dept\\
        \hline
        Alice & A \\
        Alice & B\\
        \hline
      \end{tabular}
      \hspace{20pt}
      \begin{tabular}{|c|c|c|}
        \multicolumn{2}{c}{$Mgr$}\\
        \hline
        Dept & Name & T\\
        \hline
        A & Mary & 2\\
        B & Bob  & 1\\
        B & Mary & 3\\
        \hline
      \end{tabular}
    \end{center}
    \caption{\label{tab:tab4} Instance $I_0$}
  \end{table}

  \noindent 
  Assume that we have two functional dependencies 
  $Emp:Name\rightarrow Dept$ and $Mgr:Dept\rightarrow Name$. This
  database contains two conflicts: 1) in relation $Emp$ between the
  tuples $(Alice,A)$ and $(Alice,B)$; 2) in relation $Mgr$ between the
  tuples $(B,Mary,3)$ and $(B,Bob,1)$ (Note that one person
  can be the manager of more than one department). Each of those
  conflicts can be resolved in two different ways by assuming that
  one tuple is correct and removing the other. This leads to four
  different repairs:
  \small 
  \begin{align*}
  I_1 &=\{Emp(Alice,A), Mgr(A,Mary,2), Mgr(B,Bob,1) \},\\
  I_2 &=\{Emp(Alice,B), Mgr(A,Mary,2), Mgr(B,Bob,1) \},\\
  I_3 &=\{Emp(Alice,A), Mgr(A,Mary,2), Mgr(B,Mary,3) \},\\
  I_4 &=\{Emp(Alice,B), Mgr(A,Mary,2), Mgr(B,Mary,3) \}.
  \end{align*}
  \normalsize
  For example, the repair $I_1$ is obtained by assuming that $Alice$
  works in department $A$ and the manager of department $B$ is $Bob$. 
  Since in every repair $Mary$ is the manager of the department $A$,
  we can infer that {\em true} is the consistent answer to the query 
  \[
  \phi_1 = Mgr(A,Mary).
  \]
  However it is not certain that $Alice$ works in a department managed  
  by $Mary$, i.e. {\em true} is not the consistent answer to the
  following query  
  \[
  \phi_2 = \exists x. Emp(Alice,x) \land Mgr(x,Mary).
  \]
  This is because of the repair $I_2$, where $\phi_2$ is false. 
\end{example}

As it is shown in the previous example, each conflict can be
resolved in two different ways. The framework of \cite{ArBeCh99} does
not provide any means to favor one way over another. However, in many
cases some additional information which can be used to provide a
resolution of some conflicts is available. For example:  
\begin{itemize}
  \itemsep 0pt
\item In e-commerce applications, data are accompanied with
  the timestamp of creation/last modification --- the conflicts can be
  resolved by removing from consideration old, outdated tuples.
\item In data integration scenarios, it is often possible to provide
  a (partial) order on the sources, capturing the reliability of
  contributed information --- the most reliable data can be used to
  resolve conflicts.
\item Statistics can be used to resolve conflicts created by
  misspellings.
\end{itemize}
\begin{example}[cont. Example \ref{ex:cqa}]
  Suppose that the column $T$ of the table $Mgr$ contains for each
  tuple its creation timestamp (lower values correspond to older
  tuples). We can use this information to express the preference that
  if some tuples of $Mgr$ are conflicting, the older should be removed
  from consideration (but not removed from the database). Since the
  tuple $(B,Bob,1)$ is older than $(B,Mary,3)$, we consider only the
  repairs containing the latter one: $I_3$ and $I_4$. In such a case
  we can also infer that it is certain that $Alice$ works in the
  department managed by $Mary$, i.e. {\em true} is the {\em preferred}
  consistent answer to the query $\phi_2$. 
\end{example}

In this paper we extend the framework of consistent query answers with
an additional input consisting of preference information $\Phi$. We
use $\Phi$ to define the set of {\em preferred} repairs
$\Rep^\Phi$. When we compute consistent answers, instead of considering
the set of all repairs $\Rep$, we use the set of preferred
repairs. We assume that there exists a (possibly partial) operation of
extending $\Phi$ with some additional preference information and we
write $\Phi \subseteq \Psi$ when $\Psi$ is an {\em extension} of
$\Phi$. We consider $\Phi$ to be {\em maximal} when it cannot be
extended further. The main objective of our research is to develop a
framework of preferred repairs that fulfills the following postulates: 
\begin{enumerate}
  \itemsep 0pt
  \item {\bf Non-emptiness}
    \begin{equation}
      \Rep^\Phi \neq \varnothing. \tag{$\mathcal{P}1$} \label{eq:p1}
    \end{equation}
  \item {\bf Non-discrimination}: if no preference information is
    given, then no repair is removed from consideration
    \begin{equation}
      \Rep^\varnothing = \Rep. \tag{$\mathcal{P}2$} \label{eq:p3}
    \end{equation}
  \item {\bf Monotonicity}: extending preferences can only narrow
    the set of preferred repairs
    \begin{equation}
     \Phi \subseteq \Psi \Rightarrow
     \Rep^\Psi \subseteq \Rep^\Phi. \tag{$\mathcal{P}3$} \label{eq:p4}
    \end{equation}
  \item {\bf Categoricity}: given maximal preference information
    we obtain exactly one repair
    \begin{equation}
      \text{$\Phi$ is maximal} \Rightarrow |\Rep^\Phi| = 1.
      \tag{$\mathcal{P}4$} \label{eq:p5}
    \end{equation}
\end{enumerate}
We note here that the postulates \ref{eq:p1} and \ref{eq:p3} together
imply an important property of {\bf conservativeness}: preferred
repairs are a subset of the standard repairs.

Another important goal of our research is to determine the
computational implications of introducing preferences. For this
purpose we study here two fundamental decision problems in
inconsistent databases \cite{ChMa04}:
(i) {\em repair checking} --- finding if a given database is a
preferred repair;
(ii) {\em computing consistent answers} ---  finding if an answer to a
query is present in every preferred repair.

The main contributions of this paper are:
\begin{itemize}
  \itemsep 0pt 
  \item A general and intuitive framework for incorporating preferences
  into inconsistency handling based on the notion of priority. 
  \item A study of the semantic and computational properties of two
  instantiations of the framework: (locally preferred) \l-repairs and
  (globally preferred) \g-repairs. 
\end{itemize}

\section{Basic notions and definitions}
\label{sec:basic}
In this paper, we work with databases over a schema consisting of only
one relation $R$ with attributes from $U$. We use $A,B,\ldots{}$ to
denote elements of $U$ and $X,Y,\ldots{}$ to denote subsets of
$U$. We consider two disjoint domains: uninterpreted names $D$ and
natural numbers $N$. Every attribute in $U$ is typed. We assume that
constants with different names are different and that symbols $=$,
$\neq$, $<$, $>$ have the natural interpretation over $N$. 

The instances of $R$, denoted by $r,r',\ldots{}$, can be seen as
finite, first-order structures, that share the
domains $D$ and $N$. For any tuple $t$ from $r$ by $t.A$ we denote the
value associated with the attribute $A$. In this paper we consider
first-order queries over the alphabet consisting of $R$ and binary
relation symbols $=$, $\neq$, $<$, and $>$. 

The limitation to only one relation is made only for the sake of
clarity and along the lines of \cite{ChMaSt04} the framework can be 
easily extended to handle databases with multiple relations. 

\subsection{Inconsistency and repairs}
The class of integrity constraints we study consists of functional
dependencies. 
We use $X \rightarrow Y$ to denote the following constraint:
\[
\forall t_1,t_2 \in R. \bigwedge_{A \in X} t_1.A = t_2.A
\Rightarrow \bigwedge_{B \in Y} t_1.B = t_2.B;
\]
We use this formula to identify tuples creating conflicts.
\begin{definition}[Conflicting tuples]
  Given a set of functional dependencies $F$, two tuples $t_1,t_2$
  are {\em conflicting} w.r.t $F$, denoted $t_1\conflicts_F t_2$, if
  and only if there exists a functional dependency 
  $X\rightarrow Y\in F$ such that $t_1.A = t_2.A$ for all $A \in X$ and 
  $t_1.B \neq t_2.B$ for some $B \in Y$. 
\end{definition}
\begin{definition}[Inconsistent database]
  A database $r$ is {\em inconsistent} with a set of constraints $F$ if
  and only if $r$ contains some conflicting tuples. Otherwise, the
  database is {\em consistent}.  
\end{definition}
In the general framework when repairing a database we consider two
operations: adding or removing a tuple. Because in the presence of
functional dependencies adding new tuples cannot remove conflicts, we
only consider repairs obtained by deleting tuples from the original  
instance. 
\begin{definition}[Repair]\label{def:repair}
  Given a database $r$ and a set of integrity constraints $F$, a
  database $r'$ is a {\em repair} of $r$ w.r.t. $F$ if $r'$ is a
  maximal  subset of $r$ consistent with $F$.

  \noindent
  We denote by $\Rep_F(r)$ the set of all repairs of $r$ w.r.t $F$.
\end{definition}
A repair can be viewed as the result of a process of cleaning the
input relation.  Note that since every conflict can be resolved in
two different ways and conflict are often independent, there may be an
exponential number of repairs. Also, the set of repairs of a
consistent relation $r$ contains only $r$.
\subsubsection{Conflict graphs}

\begin{definition}[Conflict graph] {\small \cite{ABCHRS03}}
  A {\em conflict graph} $G_{r,F}$ is a graph whose set of vertices is
  equal to $r$ and two tuples $t_1,t_2$ are adjacent only if they are
  conflicting (i.e. $t_1\conflicts_F t_2$).
\end{definition}
Recall that a maximal independent set of a graph $G$ is a maximal set
of vertices that contains no edge from $G$. By $\MIS(G)$ we denote the
set of all maximal independent sets of $G$. The following observation
explains why the conflict graph is considered a
{\em compact representation of all repairs}.
\begin{fact}
  For any database  $r$ and any set of functional dependencies
  $F$ we have that
  \[
  \Rep_F(r) = \MIS(G_{r,F}).
  \]
\end{fact}
\subsection{Priorities and preferred repairs}
For the clarity of presentation we assume that from now on we work
with a fixed database instance $r$ and a fixed set of  functional
dependencies $F$.

To represent the preference information, we use (possibly partial)
orientations of the conflict graph. It allows us to express 
preferences at the level of single conflicts. 
\begin{definition}[Priority]
  A binary relation $\prec \subseteq r \times r$ is a {\em priority}
  if: 
  \begin{enumerate}
    \itemsep0pt
  \item $\prec$ is asymmetric, i.e.
    \[
    \forall x,y \in r. \neg [x \prec y \land y \prec x],
    \]
  \item $\prec$ is defined only on conflicting tuples, i.e.
    \[
    \forall x,y \in r . x \prec y\Rightarrow x \conflicts_F y.
    \]
  \end{enumerate}
  If $x \prec y$ we say that the pair $\{x,y\}$ is {\em prioritized}
  and that $y$ {\em dominates} over $x$.
  A priority $\prec$ is {\em total} if every pair of
  conflicting tuples is prioritized by $\prec$.
  A priority $\prec$ is {\em acyclic} if there does not
  exist $x\in r$ such that $x \prec^* x$, where $\prec^*$ is the
  transitive closure of $\prec$.
\end{definition}
The first condition of priority demands the preference information
to be unambiguous for a single conflict. The second condition ensures
that we are given only the relevant preference information. If the
second condition is not fulfilled, then it can be easily enforced by
intersecting $\prec$ with $\conflicts_F$. 

This form of preference information allows us to easily define the
the preference extension: we orient some conflicting edges that were
not oriented before.
\begin{definition}[Priority extension]\label{def:extension}
A priority $\prec'$ is an {\em extension} of a priority $\prec$ if
$\prec'$ agrees with $\prec$ where $\prec$ is defined
(i.e. $\mathord\prec'\supseteq\mathord\prec$).
%
\end{definition}
Note that $\prec$ cannot be extended further only if $\prec$ is
total. Also an extension $\prec'$ of a priority $\prec$ is also a
priority and therefore $\prec'$ is antisymmetric and defined only on
pairs of conflicting tuples. 

Now we present two methods of using a priority to restrict the set of 
all repairs of a given relation. The first one, \l-repairs, uses the 
priority to restrict the ways of constructing a repair (cleaning the
database). The process consists of multiple iterative steps and in
each of them only a limited number of conflicts is considered. The use
of the priority has a local character because the subset of priority
used in one step is not used in any further steps. The second method,
\g-repairs, uses the priority in a global fashion by selecting most
preferred repairs according to an order induced by the priority.

\subsubsection{Locally preferred repairs}
Recall a general nondeterministic procedure for constructing a maximal
independent set of a graph: as long as the graph is not empty, we 
{\em choose} a vertex, add it to the constructed set, and remove the
vertex and all its neighbors from the graph.
Depending on the choices of vertices we make, we can construct any
maximal independent set of the input graph.
Now, let's look at this procedure from the point of constructing a
repair. Each choice of a vertex corresponds to taking a single repair 
action: keeping the corresponding tuple in the relation and removing
all tuples conflicting with it. 

Since the choice of the tuple to keep is unconstrained,
every conflict can be resolved in several different ways.
We use the priority to restrict the possible ways of 
choosing the tuple that will be kept and whose conflicts will be
resolved. The chosen tuple is among those that are not dominated
at the given step of the repairing process. We use the
{\em winnow operator} \cite{ChTODS03} to formally describe the set of
tuples that we choose from:
\[
\omega_\prec(s)=\{t\in s |\neg\exists t'\in s.t\prec t'\}.
\]
Algorithm~\ref{alg:alg1} implements the construction of preferred
repairs. An {\em \l-repair} (or a {\em locally preferred} repair) is any
instance $r'$ we can obtain with this Algorithm. We denote the set of
all \l-repairs of $r$ w.r.t. $F$ and $\prec$ by $\LRep_F^\prec(r)$.
\begin{algorithm}[ht] 
\begin{algorithmic}[1]
\State{$r' \gets \varnothing$}
\State{$s \gets r$}
\While{$\omega_\prec(s) \neq \varnothing$} \label{step:stop_cond}
\State{choose any $x \in \omega_\prec(s)$}
\label{step:choice}
\State{$r'\gets r' \cup \{x\}$}
\State{$s \gets s \setminus v(x)$}
\Comment{ where $v(x) = \{x\}\cup\{y | x \conflicts_F y\} $}
\EndWhile
\State{\Return{$r'$}}
\end{algorithmic}
\caption{\label{alg:alg1} Nondeterministic construction of an \l-repair}
\end{algorithm}
Note that an \l-repair can be characterized by the sequence 
of choices made in the step \ref{step:choice} in
Algorithm~\ref{alg:alg1} (however there can be more than one such
sequence). This observation allows us to state an alternative
definition of an \l-repair.
\begin{proposition}
  \label{prop:proc_computation_path}
  Given a priority $\prec$, a  set of tuples $X$ is an \l-repair, if
  and only if there exists an ordering $x_1,\ldots{},x_n$ of $X$ such
  that for every $i\in \{0,\ldots{},n-1\}$ the following set is
  non-empty 
  \[
  ( X \setminus \{x_1,\ldots{},x_i\}) \cap
  \omega_\prec\left(r \setminus
  \big(v(x_1) \cup \ldots \cup v(x_i)\big)\right)
  \]
  and
  $\omega_\prec(r\setminus (v(x_1)\cup\ldots\cup v(x_n))=\varnothing$.
\end{proposition}

\subsubsection{Globally preferred repairs}
The next construction uses the priority directly to compare two
repairs. Intuitively, one repair is better than another if all 
the differences between them are justified by the priority.
Formally, we define \g-repairs in the following way.
\begin{definition}[Globally preferred repair]
  \label{def:decl_repair}
  Given a priority $\prec$ and two repairs $r_1,r_2\in \Rep_F(r)$, we say
  that $r_2$ is {\em preferred} over $r_1$, and write $r_1 \ll r_2$, if
  \[
  \forall x \in r_1 \setminus r_2 . \; 
  \exists y \in r_2 \setminus r_1 . \; x \prec y.
  \]
  A repair is a \g-repair (or a {\em globally} preferred repair) if it
  is a $\ll$-maximal repair. By $\GRep_F^\prec(r)$ we denote the set
  of all \g-repairs. 
\end{definition}
This particular ``lifting'' of a preference on objects to a preference
on sets of objects can be found in other contexts. For example, a
similar definition is used for a preference among different models of
a logic program \cite{NiVe02}, or for  a preference among different
worlds \cite{Ha97}.
\subsection{Consistent query answers}
In this paper, we use a generalized notion of consistent query answers.
Instead of taking the set of all repairs, as in \cite{ArBeCh99}, we
consider families of preferred repairs. 
We only study closed first-order logic queries. We can easily
generalize our approach to open queries along the lines of
\cite{ArBeCh99, ChMaSt04}. For a given query $\varphi$ we say that
{\em true} is an answer to $\varphi$ in $r$, if $r \models \varphi$ 
in the standard model-theoretic sense.  
\begin{definition}[$\mathcal{H}$-Consistent query answer] 
Given a closed query $\varphi$ and a family of repairs 
$\mathcal{H} \subseteq \Rep_F(r)$, {\em true} is the
{\em $\mathcal{H}$-consistent query answer} to a query $\varphi$
if for every repair $r'\in\mathcal{H}$ we have $r' \models \varphi$.
\end{definition}
Note that we obtain the original notion of consistent query answer
\cite{ArBeCh99} if we take for $\mathcal{H}$ the whole set of repairs
$\Rep_F(r)$. 

In this paper, we study the cases when we take for $\mathcal{H}$
either the set of \l-repairs or the set of \g-repairs. This gives us
two notions: 
\begin{enumerate}
\itemsep0pt
\item {\em \l-preferred consistent query answer} if 
  $\mathcal{H}=\LRep_F^\prec(r)$,
\item {\em \g-preferred consistent query answer} if 
  $\mathcal{H}=\GRep_F^\prec(r)$.
\end{enumerate}
We write $r \models_{F,\prec}^\l \phi$ ($r\models_{F,\prec}^\g\phi$) 
to denote that {\em true} is the \l-preferred (resp. \g-preferred) consistent
answer to $\varphi$ (in $r$ w.r.t. $F$ and $\prec$).

\section{Basic properties}
\label{sec:properties}
\subsection{Cyclic priorities}
Before discussing specific properties of preferred repairs, we present
reasons for removing cyclic priorities from consideration. 
\begin{example}\label{ex:ex1}
  Assume a database schema $R(A,B)$ and a set of functional
  dependencies $F=\{A\rightarrow B, B\rightarrow A\}$. Consider the
  following database
  \[
  r=\{t_a=(1,1),t_b=(1,2),t_c=(2,2),t_d=(2,1)\}
  \]
  and a total cyclic priority
  $\prec=\{(t_a,t_b),(t_b,t_c),(t_c,t_d),(t_d,t_a)\}$.
  The set of all repairs is
  \[
  \Rep_F(r) = \{r_1 = \{t_a,t_c\}, r_2 = \{t_b,t_d\}\}.
  \]
  As we can easily find $\LRep_F^\prec(r)$ is empty. It is also easy to
  see that $r_1 \ll r_2$ and $r_2 \ll r_1$ and thus 
  $\GRep_F^\prec = \varnothing$. This violates the postulates
  \ref{eq:p1} and \ref{eq:p5}.
\end{example}
Intuitively, a cycle in the conflict graph represents a mutually
dependent group of conflicts (a solution of one conflict may restrict
the ways of solving other conflicts). Our intention is to break the
cycle by choosing a $\prec$-maximal element. If $\prec$ is cyclic,
then such element does not exist, which makes the construction of a
preferred repair impossible. We find this kind of preference
information (cyclic priority) to be incoherent and we exclude it form
our considerations.  

\subsection{Order properties of $\ll$}
When we restrict our considerations only to acyclic priorities,
the relation $\ll$ has interesting order properties. 
\begin{proposition} \label{prop:decl_order}
If $\prec$ is an acyclic priority and the binary relation $\ll$ on
$Rep_F(r)$ is defined in terms of $\prec$ as in Definition
~\ref{def:decl_repair}, then 
\begin{enumerate}
\itemsep 0pt
\item $\ll$ is reflexive,
\item $\ll$ is anti-symmetric,
\item $\ll$ is transitive, provided that $\prec$ is transitive.
\end{enumerate}
\end{proposition}

\begin{proof}
Before proving the main thesis we will introduce one definition and
show its two properties
\begin{definition}[Alternating chain]
Given two sets $A,B\subseteq r$ and a priority $\prec$,  
an {\em $(A,B)$-alternating $\prec$-chain} is a 
(possibly infinite) sequence $\alpha_1,\alpha_2,\ldots{}$ such that: 
\begin{itemize}
\itemsep 0pt
\item every element with even index belongs to $A$
\[
\alpha_{2*i} \in A
\]
\item every element with odd index belongs to $B$
\[
\alpha_{2*i+1} \in B
\]
\item $\prec$ holds between every two consecutive elements, i.e.
\[
\alpha_i \prec \alpha_{i+1}
\]
\end{itemize}
We say that an $(A,B)$-alternating $\prec$-chain is {\em maximal} if
it's not a proper prefix of some $(A,B)$-alternating 
$\prec$-chain\footnote{A sequence $\{a_i\}_{i=1}^n$ is a proper prefix
of a sequence $\{b_i\}_{i=1}^m$ if and only if $n<m$ and $a_i=b_i$ for
every $i\in\{1,\ldots{},n\}$. Note that $\{b_i\}_{i=1}^m$ can be
infinite ($m=\infty$), but an infinite sequence cannot have a proper
prefix.}. 
\end{definition}
When $\prec$ will be know from the context instead of saying that 
$\{\alpha_i\}$ is an $(A,B)$-alternating $\prec$-chain we will simply
say that $\{\alpha_i\}$ is an $(A,B)$-chain. 
\begin{proposition}\label{prop:inf_seq}
For any acyclic priority $\prec$ and any two sets $A,B\subseteq r$
every $(A,B)$-chain is finite.
\end{proposition}
\begin{proof}
Suppose there exists such an infinite $(A,B)$-chain  
$\{\alpha_i\}$. Because $r$ is finite, $\{\alpha_i\}$ contains a
recurrent element $x$. Thus
\[
x \prec \ldots{} \prec x.
\]
This gives us a contradiction with $\prec$ being acyclic. 
\end{proof}

\begin{proposition}\label{prop:altll}
For any acyclic priority $\prec$, and any two sets $X,Y \subseteq r$
such that $X\ll Y$ (where $\ll$ is defined in terms of $\prec$), any
maximal $(X\setminus Y,Y\setminus X)$-chain is of even length (it ends
with an element from $Y\setminus X$).  
\end{proposition}

\begin{proof}
By previous proposition we have that any 
$(X\setminus Y,Y\setminus X)$-chain is finite. Assume now that, there   
exists a maximal $(X\setminus Y,Y\setminus X)$-chain of odd length
(i.e. ending with an element from $X\setminus Y$):
\begin{equation}\label{eq:seq1}
x_1 \prec y_1 \prec x_2 \prec y_2 \prec \ldots{} \prec x_k.
\end{equation}
Since $X\ll Y$, there exists $y_k \in Y\setminus X$ such that 
$x_k\prec y_k$. Thus \eqref{eq:seq1} is a prefix of the following
$(X\setminus Y,Y\setminus X)$-chain:
\[
x_1 \prec y_1 \prec x_2 \prec y_2 \prec \ldots{} \prec x_k \prec y_k.
\]
This contradicts the maximality of \eqref{eq:seq1}.
\end{proof}

\noindent
We also state a trivial fact
\begin{fact}\label{fact:altstart}
For any acyclic priority $\prec$, any two sets $X,Y\subseteq r$ such
that $X\ll Y$, and any $x\in X \setminus Y$ there exists an
$(X\setminus Y,Y\setminus X)$-chain that starts
with $x$.
\end{fact}
Now, we show the order properties of $\ll$:
\begin{enumerate}
\itemsep 0pt
\item[1.] $\ll$ is reflexive.

  Because universal quantification over empty set is
  true, then trivially $X\ll X$ for any set $X\subseteq r$.
\item[2.] $\ll$ is asymmetric.

  Take two different sets $X,Y \subseteq r$ such that 
  $X\ll Y$ and $X \ll Y$, i.e.:
  \begin{gather} 
  \forall x \in X\setminus Y. \exists y \in Y \setminus X. x \prec y, 
  \label{eq:genY}\\
  \forall y \in Y\setminus X. \exists x \in X \setminus Y. y \prec x.
  \label{eq:genX}
  \end{gather}
  W.l.o.g we can assume that $X\setminus Y \neq \varnothing$. Take any 
  $x_1 \in X\setminus Y$. By \eqref{eq:genY} we are able to find 
  $y_1\in Y  \setminus X$ such that $x_1 \prec y_1$. Now, by
  \eqref{eq:genX} we are able to find $x_2\in X\setminus Y$ such that
  $y_1 \prec x_2$. This way we can construct an infinite 
  $(X \setminus Y,Y \setminus X)$-chain.
  This contradicts Proposition~\ref{prop:inf_seq}.
\item[3.] If $\prec$ is transitive, then $\ll$ is transitive.

  Assume $\prec$ is transitive and take three different sets
  $X,Y,Z\subseteq r$ such that $X\ll Y$ and $Y\ll Z$ (the case when two
  sets are equal is trivial). Note that:
  \begin{gather}
  \forall x\in X\setminus Y. \exists y \in Y\setminus X. x \prec y,
  \label{eq:xlly}\\
  \forall y\in Y\setminus Z. \exists z \in Z\setminus Y. y \prec z.
  \label{eq:yllz}
  \end{gather}
  Now we take any $x\in X\setminus Z$ and consider two cases depending
  if $x\in Y$ or not. 

  Suppose $x \in Y$. Let $x \prec{} \ldots {}\prec z$ be a maximal 
  $(Y\setminus Z,Z\setminus Y)$-chain where $z \in Z \setminus Y$. 
  (the existence of such a chain is by Proposition \ref{prop:altll} and
  Fact \ref{fact:altstart}). If there exists an element $z'$ of this
  chain that belongs to $Z \setminus X$ then by transitivity of
  $\prec$ we have $x \prec z'$ (which end this path of the proof). 
  Suppose that none of the elements of the 
  $(Y\setminus Z,Z\setminus Y)$-chain belongs to $Z \setminus X$, then
  in particular $z$ belongs to $X \setminus Y$. By
  \eqref{eq:xlly} there exists $y \in Y\setminus X$ such that
  $z \prec y$. Moreover $y\in Z$ or otherwise we get a contradiction
  of the maximality of the $(Y\setminus Z,Z\setminus Y)$-chain. By
  transitivity of $\prec$ we get $x\prec y$ and obviously 
  $y\in Z\setminus X$;  

  Similarly we deal with the case when $x \not \in Y$. Take 
  $x \prec{} \ldots{} {}\prec y$ to be a maximal 
  $(X\setminus Y, Y\setminus X)$-chain, where $y\in Y \setminus X$. If
  there exists an element $z'$ of this sequence that belongs to 
  $Z \setminus X$, then by transitivity of $\prec$ we have 
  $x\prec z'$ (which end this path of the proof). Suppose that none of
  the elements of the $(X\setminus Y,Y\setminus X)$-chain belongs to 
  $Z \setminus X$, then in particular $y$ belongs to $Y \setminus Z$. 
  By \eqref{eq:yllz} there exists $z\in Z\setminus Y$ such that $y\prec z$. 
  Moreover $z \not \in X$ or otherwise we get a contradiction of the
  maximality of the $(X\setminus Y,Y\setminus X)$-chain. Finally, by
  transitivity of $\prec$ we get $x\prec z$ and obviously 
  $z \in Z\setminus X$. This ends the proof.
\end{enumerate}
\end{proof}

\noindent
The following example shows that $\ll$ may not be transitive if the
underlying priority is not transitive.
\begin{example}
Consider a database 
\[
r= \{t_a=(1,1), t_b=(1,2), t_c=(1,3)\}
\]
over the schema $R(A,B)$ with one functional dependency 
$F=\{A\rightarrow B\}$ and with priority 
$\mathord\prec=\{(t_a,t_b),(t_b,t_c)\}$.
There are three repairs of $r$:
\[
\Rep_F(r) = \{A=\{t_a\}, B=\{t_b\}, C=\{t_c\}\}
\]
The corresponding conflict graph is presented on Figure
\ref{fig:fig6}. 
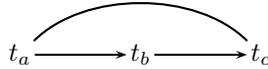
\begin{figure}[ht] 
  \begin{center}
    \begin{pspicture}(3.4,1.6)
      \rput(0.2,.5){\rnode{Ta}{$t_a$}}
      \rput(1.8,.5){\rnode{Tb}{$t_b$}}
      \rput(3.4,.5){\rnode{Tc}{$t_c$}}
      \ncline[nodesep=1.4pt]{->}{Ta}{Tb}
      \ncline[nodesep=1.4pt]{->}{Tb}{Tc}
      \nccurve[angleA=45,angleB=135,nodesep=1.4pt]{-}{Ta}{Tc}      
    \end{pspicture}
  \end{center}
  \caption{\label{fig:fig6} Conflict graph $G_{r,F}$ with orientation
    $\prec$}
\end{figure}
We note that $A\ll B$ and $B \ll C$ but $A \nll C$.
\end{example}

\subsection{Fulfillment of the postulates}
Before we prove the fulfillment of the postulates
\ref{eq:p1}--\ref{eq:p5} we state an important property of the two
instantiations of preferred repairs: constructing a repair from the
locally best tuples by the notion of \l-repairs conforms with the
global notion of preference
(\g-repairs). \begin{theorem}\label{thm:proc_subsumes_decl}  
  If $\prec$ is an acyclic priority, then
  \[
  \LRep_F^\prec(r) \subseteq \GRep_F^\prec(r).
  \]
\end{theorem}
\begin{proof}
Induction over the size of $r$. Trivial for $r=\varnothing$.

Assume the hypothesis holds for any proper subset of $r$ and there
exists $X\in  \LRep_F^\prec(r)$ such that $X\ll Y$ for some
$Y\in \Rep_F(r)$. By Proposition \ref{prop:proc_computation_path}
$\omega_\prec(r) \cap X$ is non-empty. Take then any 
$x\in \omega_\prec(r)\cap X$. $x\in Y$ or otherwise we receive a
contradiction $X\ll Y$. Note that $Y\setminus\{x\}$ is a repair of
$r\setminus v(x)$ and $X \setminus \{x\}$ even a \l-repair of
$r\setminus v(x)$. Moreover $X \setminus \{x\} \ll Y\setminus \{x\}$
in terms of the database $r\setminus v(x)$. Thus $X\setminus \{x\}$ is
not \g-repair of $r\setminus\{x\}$, which is a contradiction of the
inductive hypothesis.  
\end{proof}

\noindent
In the following example we observe that the reverse containment
does not hold for an arbitrary acyclic priority, i.e. the construction
of \l-repairs by choosing only the best elements locally (as in \l-repairs)
may miss a \g-repair.
\begin{example}\label{ex:proc_diff_decl}
  Consider a database  
  \[
  r = \{t_a=(1,1,1),t_b=(2,1,2),t_c=(3,1,3),t_d=(4,1,3)\}
  \]
  over the  schema $R(A,B,C)$ with a set of functional dependencies
  $F=\{B\rightarrow C\}$ and a acyclic priority
  \[
    \mathord{\prec}=\{(t_c,t_a),(t_d,t_b)\}
  \]
  The set of repairs is
  $\Rep_F(r) = \{r_1 = \{t_a\}, r_2 = \{t_b\}, r_3 = \{t_c,t_d\}\}$.
  As we can easily find $\GRep_F^\prec(r)=\Rep_F(r)$. 
  Because each of the $t_c$ and $t_d$ is dominated, the \g-repair 
  $r_3$ is not an \l-repair, and thus $\LRep_F^\prec(r) = \{r_1,r_2\}$. 
\end{example}
Later on we present sufficient conditions under which both
instantiations of preferred repairs are equivalent 
(Theorem \ref{thm:proc_equals_decl}).

We recall that extending priority consists of prioritizing conflicts
not prioritized before and a priority that cannot be extended further
(i.e. is maximal) is a total priority. 
Both classes of referred repairs that we consider satisfy the
postulates \ref{eq:p1} -- \ref{eq:p5}:
\begin{theorem}[\ref{eq:p1}--\ref{eq:p5} for $\LRep$]
  \label{thm:proc_postulates}
  For every relation instance $r$, set of functional dependencies $F$,
  and acyclic priority $\prec$, $\LRep_F^\prec(r)$ satisfies 
  \ref{eq:p1}--\ref{eq:p5}.
\end{theorem}
\begin{proof}
We receive \ref{eq:p1} from the fact that if $\prec$ is acyclic then
$\omega_\prec(X)$ is non-empty if and only if $X$ is non-empty.

\ref{eq:p3} is implied by the fact that $\omega_\varnothing$ is an
identity function what makes $\LRep$ a generic procedure for
constructing all maximal independent sets of $G_{r,C}$.

To prove \ref{eq:p4} assume that $\prec',\prec$ are acyclic priorities
such that $\prec'\subseteq\prec$. Take then any $X\in \LRep_F^\prec(r)$
and let $\sigma$ be any ordering of $X$ from Proposition
\ref{prop:proc_computation_path}. Note that since for any set $A$ we
have $\omega_\prec(A) \subseteq \omega_{\prec'}(A)$ then $\sigma$ also
fulfills conditions of Proposition \ref{prop:proc_computation_path} in
terms of $\prec'$.

\ref{eq:p5} is a consequence of \ref{eq:p1} for $\LRep$, Theorem
\ref{thm:proc_subsumes_decl}, and \ref{eq:p5} for $\GRep$. 
\end{proof}

\begin{theorem}[\ref{eq:p1}--\ref{eq:p5} for $\GRep$]
  \label{thm:decl_postulates}
  For every relation instance $r$, set of functional dependencies $F$,
  and acyclic priority $\prec$, $\GRep_F^\prec(r)$ satisfies 
  \ref{eq:p1}--\ref{eq:p5}.
\end{theorem}

\begin{proof}
We get \ref{eq:p1} from the definition.

With an empty priority we cannot justify $X \ll Y$ for any two
different repairs $X$ and $Y$, what implies \ref{eq:p3}.

To show \ref{eq:p4} assume that $\prec',\prec$ are acyclic priorities
such that $\prec'\subseteq\prec$, $X\in \GRep_F^\prec(r)$, and suppose
there exists $Y\in \GRep_F^{\prec'}(r)$ such that $Y$ is preferred over
$X$ in terms of $\prec'$. But since $\prec'\subseteq \prec$ this
implies that $Y$ is also preferred over $X$ in terms of $\prec$. This
is a contradiction. 

In order to prove \ref{eq:p5} assume there exist two different
repairs $X$ and $Y$ in $\GRep_F^\prec(r)$. $X \not \ll Y$ implies that
there exists an element $x \in X \setminus Y$ such that for any
conflicting with $x$ tuple $y$ from $Y\setminus X$ we have 
$x \not \prec y$. Since $\prec$ is total for any such $y$ we have 
$y \prec x$. Take all such tuples $y_1,\ldots{},y_n$ and by $Y'$ denote
any repair that contains the following elements
\[
Y\setminus \{y_1,\ldots{},y_n\} \cup \{x\}
\]
Such a repair exists because this set contains no conflicting
tuples. Obviously $Y'\neq Y$ and at the same time $Y \ll Y'$. This
contradicts that $Y\in \GRep_F^\prec(r)$.
\end{proof}

\subsection{Equivalence of $\LRep$ and $\GRep$}
As we showed in Example \ref{ex:proc_diff_decl} $\LRep$ doesn't have to be equal to
$\GRep$. It suffices, however, to remove from consideration priorities
with cyclic extensions to obtain the equivalence of the two notions of 
preferred repair:
\begin{theorem} \label{thm:proc_equals_decl}
  If $\prec$ is a priority having only acyclic extensions, then
  \[
  \GRep_F^\prec(r) =  \LRep_F^\prec(r).
  \]
\end{theorem}
\begin{proof}
We need to show $\GRep_F^\prec(r) \subseteq \LRep_F^\prec(r)$.
Take any $X\in \GRep_F^\prec(r)$ and construct $\prec'$ a total
extension of $\prec$ by prioritizing (un-prioritized by $\prec$)
conflicts in favor for $X$, i.e. $\prec'$ is any total priority such
that for any $x\in X$ and any $y$ if $x\conflicts_F y$ and 
$x \not \prec y$ then $y\prec x$. Since $\prec$ has only acyclic
extensions $\prec'$ is acyclic. It should be clear from the
construction that $X\in \GRep_F^{\prec'}(r)$.   
By \ref{eq:p1}, \ref{eq:p3}, \ref{eq:p5} and Theorem
\ref{thm:proc_subsumes_decl} this implies that $X\in
\LRep_F^{\prec'}(r)$. This by \ref{eq:p4} gives us that $X\in
\LRep_F^\prec(r)$.  
\end{proof}

\noindent
The following example shows, however, that the requirement of no
cyclic extensions is not necessary for the equality above to hold. 
\begin{example}
Consider schema $R(A,B,C)$ together with a set of functional
dependencies $F=\{B\rightarrow C\}$. Suppose we have
a database:
\[
r = \{t_a=(1,1,1),t_b=(2,1,1),t_c=(3,1,2),t_d=(4,1,2) \}
\]
with a priority $\mathord\prec=\{(t_c,t_a),(t_d,t_b)\}$. 
The conflict graph is presented on Figure \ref{fig:fig1}.
\begin{figure}[ht] 
  \begin{center}
    \begin{pspicture}(2,1.3)
      \rput(.4,1.1){\rnode{Ta}{$t_a$}}
      \rput(1.6,1.1){\rnode{Tb}{$t_b$}}
      \rput(.4,.1){\rnode{Tc}{$t_c$}}
      \rput(1.6,.1){\rnode{Td}{$t_d$}}
      \ncline[nodesep=1.4pt]{<-}{Ta}{Tc}
      \ncline[nodesep=1.4pt]{-}{Ta}{Td}
      \ncline[nodesep=1.4pt]{-}{Tb}{Tc}
      \ncline[nodesep=1.4pt]{<-}{Tb}{Td}
    \end{pspicture}
  \end{center}
  \caption{\label{fig:fig1} Conflict graph $G_{r,F}$ with orientation
    $\prec$}
\end{figure}
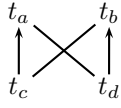
$\prec$ has a cyclic extension
$\mathord\prec'=\mathord\prec \cup \{(t_a,t_d),(t_b,t_c)\}$. At the
same time $\LRep_F^\prec(r) = \GRep_F^\prec(r) = \{\{t_a,t_b\}\}$.
\end{example}
\section{Computational properties}
\label{sec:complexity}
We study two fundamental problems of handling inconsistencies with
priorities: (i) {\em repair checking} -- determining if a
database is a preferred repair of a given database; (ii) 
{\em consistent query answers} -- checking if {\em true} is an answer
to a given query in every preferred repair. 
We use the notion of {\em data complexity} \cite{Var82} which captures
the complexity of a problem as a function of the number of tuples in
the database. The database schema, the integrity constraints, and the
query are assumed to be fixed.
\subsection{Locally preferred repairs}
Recall Algorithm~\ref{alg:alg1} and note that because the consecutive 
choices made in the step~\ref{step:choice} consist of mutually
non-conflicting tuples, the state of the computation is independent of
the order of the choices\footnote{The state of computation means the
  repair being constructed and the possible further choices.}. Given a
repair $r'$, we can ``simulate'' its 
construction by restricting the choices in the step \ref{step:choice}
to $r'\cap \omega_\prec(r)$.  The simulation succeeds if and only if
$r'$ is an \l-repair.  
\begin{theorem} \label{thm:decl_total_tractable}
  Given a fixed set of functional dependencies $F$, the set
  \[
  B^\l_F = \{(r,r',\mathord\prec) | r' \in \LRep_F^\prec(r)\}
  \]
  is in PTIME.
\end{theorem}
It is shown in \cite{ChMa04} that computing consistent answers to
conjunctive queries is co-NP-complete, but if we consider only ground
quantifier-free queries, the problem is in PTIME. On the other hand,
computing \l-preferred consistent answers turns out to be an
intractable problem even if we consider very simple, single-atom
queries. 
\begin{theorem}
  \label{thm:proc_answers_intractable}
  There exists a set of four functional dependencies $F$ and a
  quantifier-free ground query $\varphi$ (consisting of one atom only) 
  such that the set
  \[
   D^\l_{F,\varphi}=\{ (r,\mathord\prec) | 
   r \models_{F,\mathord\prec}^\l \varphi\}, 
  \]
  is co-NP-complete.
\end{theorem}
\begin{proof}
  It's easy to construct a nondeterministic Turing machine for
  $D^l_{F,\varphi}$ following informal description presented here: The
  machine uses nondeterministic transitions to compute all
  \l-preferred repairs of $r$ and for each one checks the answer to
  $\varphi$. Note that 
  \[
  r \models_{F,\prec}^l \varphi \iff
  \forall r'\in \LRep_F^\prec(r). r' \models \varphi \iff
  \neg \exists r'\in \LRep_F^\prec(r). r' \models \neg \varphi.
  \]
  This allows us to state that the constructed machine decides the
  complement of $D^l_{F,\varphi}$.

  \noindent
  Now, consider the schema $R(A_1,B_1,\ldots{},A_4,B_4)$ with the
  set of functional dependencies
  $F=\{A_1\rightarrow B_1, \ldots{}, A_4\rightarrow B_4\}$ and a
  ground query $\neg R(b)$, where the value of $b$ can be found in
  Table \ref{tab:tab2}.

  \noindent
  We show here a polynomial reduction of the complement of $3SAT$ to
  $D^l_{\neg R(b),F}$, i.e. for any boolean formula $\varphi$ in $3CNF$
  we construct a pair $(r_\varphi,\prec_\varphi)$ of a polynomial size
  in the size of $\varphi$ and such that
  \[
  (r_\varphi,\prec_\varphi) \in  D^l_{F,\neg R(b)}
  \iff \varphi \not \in 3SAT.
  \]
  Take then any formula $\varphi$ in $3CNF$ and let $n$ be the number
  of variables used in $\varphi$ and $k$ the number of conjuncts of
  $\varphi$. For simplicity we assume that:
  \begin{itemize}
    \item used variables have consecutive indexes $x_1,\ldots{},x_n$,
    \item $\varphi = c_1 \land \ldots \land c_k$
    \item each conjunct consists of exactly three literals
      $c_j = l_{j,1} \lor l_{j,2} \lor l_{j,3}$ for $(j=1,\ldots{},k)$.
  \end{itemize}
  We define two auxiliary functions  $var$ and $sgn$ on literals in the
  following fashion:
  \begin{align*}
    var(x_i) & = i, & sgn(x_i) & = 1, \\
    var(\neg x_i) & = i, & sgn(\neg x_i) & = -1.
  \end{align*}
  The constructed database contains the following elements:
  \[
  r_\varphi=\{v_1,\bar{v}_1,\ldots{},\bar{v}_n,v_n,d_1\ldots{},d_k,b\}
  \]
  whose exact values can be found in Table \ref{tab:tab2}.
  \begin{table}[htb]\normalsize
  \begin{center}
      \[
      \begin{array}{|c|c|c|c|c|c|c|c|c|}
        \hline
                  & A_1 & B_1 & A_2 & B_2 & A_3 & B_3 & A_4 & B_4 \\
        \hline
        \hline
        v_i       &  i  &  1  &  i  & -1  &  i  & -1  &  i  & -1  \\
        \hline
        \bar{v}_i &  i  &  2  &  i  &  1  &  i  &  1  &  i  &  1  \\
        \hline
        d_j       &  0  &  1  &  var(l_{j,1})  & sgn(l_{j,1}) &
        var(l_{j,2})  & sgn(l_{j,2}) &  var(l_{j,3})  & sgn(l_{j,3}) \\
        \hline
        b        &  0  &  0  &  0  &  0  &  0  &  0  &  0  &  0   \\
        \hline
      \end{array}
      \]
  \end{center}
  \caption{\label{tab:tab2} Values of tuples in $r_\varphi$}
  \end{table}
  \noindent
  The priority relation $\prec_\varphi$ is the unique minimal binary
  relation on $r_\varphi$ satisfying the following conditions:
  \begin{align*}
    d_j & \prec_\varphi v_{var(l_{j,i})},
    & &\text{for $j\in\{1,\ldots{},k\}$, $i\in\{1,2,3\}$ such that
    $sgn(l_{j,i}) = 1$,}\\
    d_j & \prec_\varphi \bar{v}_{var(l_{j,i})},
    & &\text{for $j\in\{1,\ldots{},k\}$, $i\in\{1,2,3\}$ such that
    $sgn(l_{j,i}) = -1$,}\\
    b & \prec_\varphi d_j, & &\text{for $j\in \{1,\ldots{},k\}$.}
  \end{align*}
  Note that this priority relation is acyclic. Also note that
  construction of $(r_\varphi,\prec_\varphi)$ can be implemented in
  time polynomial in the size of the of the input formula $\varphi$.
  On Figure \ref{fig:fig3} we can find a conflict graph of an instance
  received from reduction of a formula
  $\varphi =
  (\neg x_1 \lor x_2 \lor x_3) \land
  (x_3 \lor \neg x_4 \lor x_5) \land
  (\neg x_5 \lor \neg x_6 \lor x_7)$.
  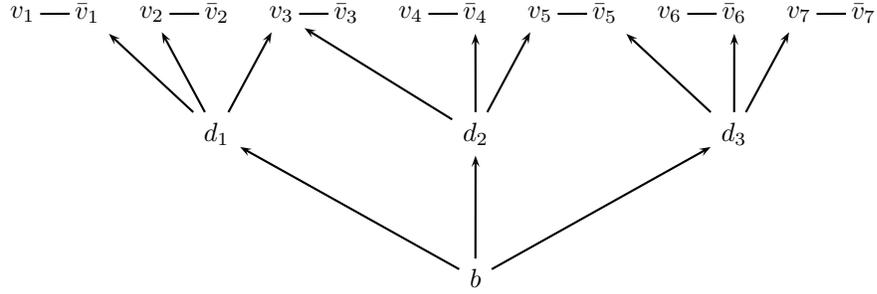
\begin{figure}[hbt]
  \begin{center}
  \psset{xunit=.86cm,yunit=.8cm}
  \begin{pspicture}(15,-6)
    \rput(1,-1){\rnode{v1}{$v_1$}}
    \rput(2,-1){\rnode{_v1}{$\bar{v}_1$}}
    \rput(3,-1){\rnode{v2}{$v_2$}}
    \rput(4,-1){\rnode{_v2}{$\bar{v}_2$}}
    \rput(5,-1){\rnode{v3}{$v_3$}}
    \rput(6,-1){\rnode{_v3}{$\bar{v}_3$}}
    \rput(7,-1){\rnode{v4}{$v_4$}}
    \rput(8,-1){\rnode{_v4}{$\bar{v}_4$}}
    \rput(9,-1){\rnode{v5}{$v_5$}}
    \rput(10,-1){\rnode{_v5}{$\bar{v}_5$}}
    \rput(11,-1){\rnode{v6}{$v_6$}}
    \rput(12,-1){\rnode{_v6}{$\bar{v}_6$}}
    \rput(13,-1){\rnode{v7}{$v_7$}}
    \rput(14,-1){\rnode{_v7}{$\bar{v}_7$}}

    \ncline[nodesep=2pt]{-}{v1}{_v1}
    \ncline[nodesep=2pt]{-}{v2}{_v2}
    \ncline[nodesep=2pt]{-}{v3}{_v3}
    \ncline[nodesep=2pt]{-}{v4}{_v4}
    \ncline[nodesep=2pt]{-}{v5}{_v5}
    \ncline[nodesep=2pt]{-}{v6}{_v6}
    \ncline[nodesep=2pt]{-}{v7}{_v7}

    \rput(4,-3){\rnode{d1}{$d_1$}}
    \rput(8,-3){\rnode{d2}{$d_2$}}
    \rput(12,-3){\rnode{d3}{$d_3$}}

    \ncline[nodesep=4pt]{<-}{_v1}{d1}
    \ncline[nodesep=4pt]{<-}{v2}{d1}
    \ncline[nodesep=4pt]{<-}{v3}{d1}
    \ncline[nodesep=4pt]{<-}{v3}{d2}
    \ncline[nodesep=4pt]{<-}{_v4}{d2}
    \ncline[nodesep=4pt]{<-}{v5}{d2}
    \ncline[nodesep=4pt]{<-}{_v5}{d3}
    \ncline[nodesep=4pt]{<-}{_v6}{d3}
    \ncline[nodesep=4pt]{<-}{v7}{d3}

    \rput(8,-5.4){\rnode{b}{$b$}}

    \ncline[nodesep=4pt]{<-}{d1}{b}
    \ncline[nodesep=4pt]{<-}{d2}{b}
    \ncline[nodesep=4pt]{<-}{d3}{b}
  \end{pspicture}
  \end{center}
  \caption{\label{fig:fig3} Conflict graph for
     $\varphi = (\neg x_1 \lor x_2 \lor x_3) \land
      (x_3 \lor \neg x_4 \lor x_5) \land
      (\neg x_5 \lor \neg x_6 \lor x_7)$ and 
      orientation $\prec_\varphi$.}
  \end{figure}

  \noindent
  Now, we show that
  \[
  \exists r' \in \LRep_F^{\prec_\varphi}(r_\varphi). b \in r'
  \iff \varphi \in 3SAT
  \]
  \begin{itemize}
  \item[\fbox{$\Rightarrow$}]
    Fist note that since $b\in r'$ then none of the tuples
    $d_1,\ldots{},d_k$ belongs to $r'$. Therefore for every
    $i\in\{1,\ldots{},n\}$ either $v_i$ or $\bar{v}_i$ belongs to
    $r'$. Thus the following is a proper definition of a boolean
    valuation:
    \[
    V(x_i) =
    \begin{cases}
      true & \text{if $v_i \in r'$}\\
      false & \text{if $\bar{v}_i \in r'$}
    \end{cases}
    \]
    Next, we show that $\varphi$ is true for $V$. Suppose otherwise,
    i.e. there exists a conjunct $c_m$ that is not true for
    $V$. W.l.o.g. we can assume that $c_m=x_1\lor\neg x_2\lor x_3$.
    This implies that $\{v_1,\bar{v}_2,v_3\} \cap r' = \varnothing$
    and thus $\bar{v}_1,v_2,\bar{v}_3 \in r'$.

    Take $t_1,\ldots{},t_n$ to be the ordering of $r'$ from
    Proposition \ref{prop:proc_computation_path}. Since no $d_j$
    tuples are present in $r'$, and the tuple $b$ is dominated by
    every $d_j$ tuple (which in turn is dominated by some $v_i$ and
    $\bar{v}_i$ tuples) then $t_n=b$. Let $s$ be the last index of
    this sequence that $t_s$ is equal to either $\bar{v}_1$, $v_2$, or
    $\bar{v}_3$. Since $d_m$ is dominated only by $v_1$, $\bar{v}_2$,
    and $v_3$ we have for any $p \geq s$
    \[
        d_m \in \omega_{\prec_\varphi}
            \big(r_\varphi \setminus ( v(t_1) \cup \ldots{} \cup
            v(t_p))\big).
    \]
    This implies that
    $\omega_{\prec_\varphi}(r_\varphi \setminus ( v(t_1) \cup \ldots{} \cup
            v(t_n))) \neq \varnothing$ which gives a contradiction.
  \item[\fbox{$\Leftarrow$}]
    Take any valuation $V$ for which $\varphi$ is true and construct
    the following set
    \[
    r'=\{b\} \cup \{v_i|V(x_i)\} \cup \{\bar{v}_i|\neg V(x_i)\}.
    \]
    First, note that $r'$ is a repair: it contains no conflicting
    tuples and for every tuple from $r_\varphi\setminus r'$ there
    exists a conflicting tuple in $r'$.

    Next, we show that $r\in \LRep_F^{\prec_\varphi}(r_\varphi)$. In
    order to prove that we note that for any subset
    $X \subseteq r' \setminus \{b\}$ we have
    \begin{equation}\label{eq:eq1}
      d_j \not \in
      \omega_{\prec_\varphi}\left(r_\varphi \setminus
      \bigcup_{x\in X} v(x)\right), \quad
      \text{for $j=1,\ldots{},k$}.
    \end{equation}
    Suppose otherwise, i.e. there exists a set
    $X \subseteq r'\setminus \{b\}$ and $m$ such that
    \[
    d_m \in \omega_{\prec_\varphi}\left(r_\varphi \setminus
      \bigcup_{x\in X} v(x)\right).
    \]
    W.l.o.g. we can assume that $c_m=x_1\lor \neg x_2 \lor x_3$.
    From the construction of $r_\varphi$ and $\prec_\varphi$ this
    implies that $\bar{v}_1,v_2,\bar{v}_3 \in X$ which is equivalent
    with $V(x_1)=false$, $V(x_2)=true$, and $V(x_3)=false$. This
    implies that $c_m$ is not true for $V$ which yields a
    contradiction with $\varphi$ being satisfied by $V$.

    The property \eqref{eq:eq1} allows us to use Proposition
    \ref{prop:proc_computation_path} (take any ordering of $r'$ with
    $b$ on the last position) to state that $r'$ is \l-preferred
    repair w.r.t $F$ and $\prec_\varphi$.
  \end{itemize}
  It should be noted here that adding just one tuple
  $b'=(0,0,0,1,0,1,0,1)$ and extending the priority with
  $b'\prec_\varphi b$ constructs a reduction of $3SAT$ to the
  complement of $D^l_{F,R(b')}$. And therefore computing
  \l-preferred consistent answers is intractable also for
  a query consisting only of one positive literal.
\end{proof}
\subsection{Globally preferred  repairs}
Unlike \l-repairs, the notion of \g-repairs, because of its global
character, cannot be captured without an essential use of
nondeterminism. 
\begin{theorem}\label{thm:decl_repair_check_intractable}
  There exists a set of five functional dependencies $F$ such that 
  the set
  \[
  B^\g_F = \{(r,r',\mathord\prec) | r' \in \GRep_F^\prec(r)\}
  \]
  is co-NP-complete.
\end{theorem}
\begin{proof}
  It's easy to construct a nondeterministic Turing machine
  $B^g_F$. The machine first checks if $r'$ is a repair; if yes the
  machine nondeterministically computes every repair and checks if any
  of them (different than $r'$) is preferred over $r'$
  w.r.t. $\prec$. This machine decides the complement of $B^g_F$.

  Now, we show that the problem co-NP-hard by reducing the
  complement of $3SAT$ to $B^g_F$. Consider the database schema
  $R(A_1,B_1,\ldots{},A_5,B_5)$ with the following set of integrity
  constraints  $F=\{A_1\rightarrow B_1,\ldots{},A_5\rightarrow B_5\}$.
  For any boolean formula $\varphi$ in $3CNF$ we construct a triple
  $(r_\varphi,X_\varphi,\prec_\varphi)$ of size polynomial in
  the size of $\varphi$ and such that
  \[
  (r_\varphi,X_\varphi,\prec_\varphi) \in  B^g_F \iff
  \varphi \not \in 3SAT.
  \]
  Moreover the reduction can be implemented in time polynomial in the
  size of $\varphi$.

  Take then any formula $\varphi$ in $3CNF$ and let $n$ be the number
  of variables used in $\varphi$ and $k$ the number of conjuncts of
  $\varphi$. For simplicity we assume that:
  \begin{itemize}
    \itemsep0pt
    \item used variables have consecutive indexes $x_1,\ldots{},x_n$,
    \item $\varphi = c_1 \land \ldots \land c_k$
    \item each conjunct consists of exactly three literals
      $c_j = l_{j,1} \lor l_{j,2} \lor l_{j,3}$ for $(j=1,\ldots{},k)$.
  \end{itemize}
  We define two auxiliary functions  $var$ and $sgn$ on literals as  follows:
  \begin{align*}
    var(x_i) & = i, & sgn(x_i) & = 1, \\
    var(\neg x_i) & = i, & sgn(\neg x_i) & = -1.
  \end{align*}
  The constructed database contains the following elements
  \[
  r_\varphi=\{
  v_1,\bar{v}_1,\ldots{},v_n,\bar{v}_n,
  w_1,\ldots{},w_n,
  d_1,\ldots{},d_k,
  s,t
  \},
  \]
  whose exact values can be found in Table \ref{tab:tab1}.
  \begin{table}[htb]\footnotesize
  \begin{center}
      \[
      \begin{array}{|c|c|c|c|c|c|c|c|c|c|c|}
        \hline
                  & A_1 & B_1 & A_2 & B_2 & A_3 & B_3 & A_4 & B_4 & A_5 & B_5 \\
        \hline
        \hline
        v_i       &  1  &  1  &  i  &  1  &  i  &  -1  & i  & -1  &  i  & -1  \\
        \hline
        \bar{v}_i &  1  &  1  &  i  &  2  &  i  &  1  &  i  &  1  &  i  &  1  \\
        \hline
        w_i       &  2  &  2  &  i  &  3  &  0  &  0  &  0  &  0  &  0  &  0  \\
        \hline
        s         &  1  &  2  &  n+1&  1  &  0  &  0  &  0  &  0  &  0  &  0  \\
        \hline
        t         &  2  &  1  &  n+1&  2  &  0  &  0  &  0  &  0  &  0  &  0  \\
        \hline
        d_j       &  2  &  2  &  0  &  0  &   var(l_{j,1}) & sgn(l_{j,1}) &
        var(l_{j,2}) & sgn(l_{j,2}) &  var(l_{j,3}) & sgn(l_{j,3})\\
        \hline
      \end{array}
      \]
  \end{center}
  \caption{\label{tab:tab1} Values of tuples in $r_\varphi$}
  \end{table}

  \noindent
  The set $X_\varphi$ consists of the following elements
  \[
  X_\varphi =\{w_1,\ldots{},w_n,d_1,\ldots{},d_n,s\}.
  \]
  It's easy to note that $X_\varphi$ is a repair of $r_\varphi$
  w.r.t. $F$. Clearly $X_\varphi \subseteq r_\varphi$, no two elements of
  $X_\varphi$ are conflicting, and for every element from the set
  $r_\varphi \setminus X_\varphi$ there exists a conflicting element from
  $X_\varphi$ ($s$ for $t$ and $w_i$ for $v_i$ or $\bar{v}_i$).

  \noindent
  The priority relation $\prec_\varphi$ is the unique minimal binary
  relation on $r_\varphi$ satisfying the following conditions:
  \begin{align*}
    s & \prec_\varphi t,\\
    w_i & \prec_\varphi v_i, & &\text{for $i\in \{1,\ldots{},n\}$,}\\
    w_i & \prec_\varphi \bar{v}_i, & &\text{for $i\in \{1,\ldots{},n\}$,}\\
    d_j & \prec_\varphi v_i, & &
    \text{if $c_j$ uses a positive literal $x_i$,}\\
    d_j & \prec_\varphi \bar{v}_i, & &
    \text{if $c_j$ uses a negative literal $\neg x_i$.}
  \end{align*}
  Note that this priority relation is acyclic. Also note that the
  triple $(r_\varphi,X_\varphi,\prec_\varphi)$ can be constructed in
  the time polynomial in the size of the formula $\varphi$. On Figure
  \ref{fig:fig2} we can find a conflict graph of the instance received
  from reduction of the formula
  $\varphi = (x_1 \lor \neg x_2 \lor x_3) 
   \land (\neg x_2 \lor \neg x_3 \lor x_4)$.
  \begin{figure}[hbt]
  \begin{center}
  \psset{xunit=.76cm,yunit=.8cm,nodesep=4pt}
  \begin{pspicture}(17,-6)
    \rput(5,-1){\rnode{t}{$t$}}
    \rput(7,-1){\rnode{v1}{$v_1$}}
    \rput(8.5,-1){\rnode{_v1}{$\bar{v}_1$}}
    \rput(9.5,-1){\rnode{v2}{$v_2$}}
    \rput(11,-1){\rnode{_v2}{$\bar{v}_2$}}
    \rput(12,-1){\rnode{v3}{$v_3$}}
    \rput(13.5,-1){\rnode{_v3}{$\bar{v}_3$}}
    \rput(14.5,-1){\rnode{v4}{$v_4$}}
    \rput(16,-1){\rnode{_v4}{$\bar{v}_4$}}

    \rput(1,-5){\rnode{d1}{$d_1$}}
    \rput(3,-5){\rnode{d2}{$d_2$}}
    \rput(5,-5){\rnode{s}{$s$}}
    \rput(7.75,-5){\rnode{w1}{$w_1$}}
    \rput(10.25,-5){\rnode{w2}{$w_2$}}
    \rput(12.75,-5){\rnode{w3}{$w_3$}}
    \rput(15.25,-5){\rnode{w4}{$w_4$}}

    \psset{nodesep=3pt,linewidth=.4pt,linecolor=blue,arrowsize=2.5pt 3.4}

    \ncline{-}{v1}{_v1}
    \ncline{-}{v2}{_v2}
    \ncline{-}{v3}{_v3}
    \ncline{-}{v4}{_v4}

    \ncline{->}{w1}{ v1}
    \ncline{->}{w1}{_v1}
    \ncline{->}{w2}{ v2}
    \ncline{->}{w2}{_v2}
    \ncline{->}{w3}{ v3}
    \ncline{->}{w3}{_v3}
    \ncline{->}{w4}{ v4}
    \ncline{->}{w4}{_v4}

    \psset{linecolor=red}

    \ncline{-}{t}{w1}
    \ncline{-}{t}{w2}
    \ncline{-}{t}{w3}
    \ncline{-}{t}{w4}
    \ncline{-}{t}{d1}
    \ncline{-}{t}{d2}

    \ncline{<-}{t}{s}

    \psset{linecolor=green}

    \ncline{-}{s}{ v1}
    \ncline{-}{s}{_v1}
    \ncline{-}{s}{ v2}
    \ncline{-}{s}{_v2}
    \ncline{-}{s}{ v3}
    \ncline{-}{s}{_v3}
    \ncline{-}{s}{ v4}
    \ncline{-}{s}{_v5}

    \psset{linecolor=black}

    \ncline{->}{d1}{v1}
    \ncline{->}{d1}{_v2}
    \ncline{->}{d1}{v3}

    \ncline{->}{d2}{_v2}
    \ncline{->}{d2}{_v3}
    \ncline{->}{d2}{v4}

    \psframe[linecolor=gray,linestyle=dotted,linewidth=.6pt,framearc=.4]
    (.5,-5.5)(16,-4.5)
    \rput*(9,-5.5){$X_\varphi$}
  \end{pspicture}
  \end{center}
  \caption{\label{fig:fig2} Conflict graph for
     $\varphi = (x_1 \lor \neg x_2 \lor x_3) \land
    (\neg x_2 \lor \neg x_3 \lor x_4)$ and orientation $\prec_\varphi$.}
  \end{figure}
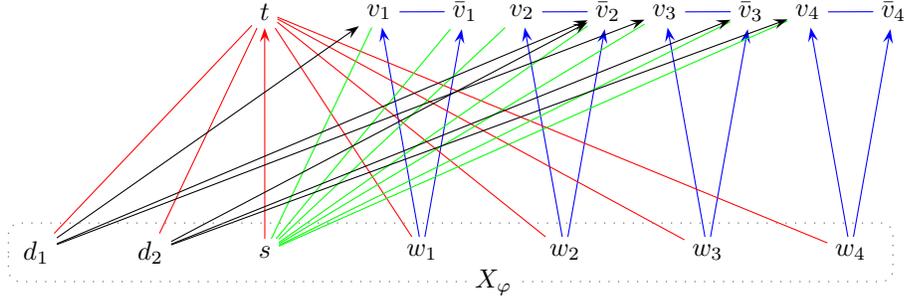

  Now, we show that for any $\varphi$ using variables
  $x_1,\ldots{},x_n$ the following holds
  \[
  X_\varphi \not \in \GRep_F^{\prec_\varphi}(r_\varphi) \iff \varphi \in 3SAT.
  \]
  \begin{enumerate}
    \item[\fbox{$\Leftarrow$}]
      Suppose $\varphi \in 3SAT$ and take
      $V:\{x_1,\ldots{},x_n\}\rightarrow\mathcal{B}$ to be the valuation
      for which $\varphi$ is true. Consider the following set
      \[
      Y_V = \{ t \} \cup \{v_i | V(x_i)\}\cup \{\bar{v}_i | \neg V(x_i)\}
      \]
      It's easy to find that $Y_V$ is a repair and moreover
      $X_\varphi \ll Y_V$. Thus $X_\varphi$ is not a maximally
      \g-preferred repair.
    \item[\fbox{$\Rightarrow$}]
      Suppose $X_\varphi \not \in \GRep_F^\prec(r)$, i.e. there exists 
      $Y \in \Rep_F(r)$ such that $X \ll Y$ and $Y \neq X$.

      First note that $t \in Y$. Otherwise for $Y$ to be preferred
      over $X$ the tuple $s$ has to be contained in $Y$ because there
      is no element dominating $s$ except for $t$. Since $s$ is
      adjacent with every $v_i$ and $\bar{v}_i$ then also none of
      $v_i$ and $\bar{v}_i$ belongs to $Y$. This implies that $Y=X$
      which is a contradiction.

      Since $t$ is adjacent to every element of $X_\varphi$ and
      $t \in Y$ the sets $Y$ and $X_\varphi$ are disjoint. This implies
      that for every $i$ the set $Y$ contains either $v_i$ or
      $\bar{v}_i$ (from maximality, independence, and the fact that
      $X \ll Y$).

      Take now the following boolean valuation
      \[
         V_Y(x_i) =
             \begin{cases}
                   true & \text{if $v_i \in Y$}\\
                  false & \text{if $\bar{v}_i \in Y$}
             \end{cases}
      \]
      We show that $V_Y$ is a valuation for which $\varphi$ is true.
      Suppose otherwise, that there exists a conjunct $c_m$ that is not
      true under $V_Y$. W.l.o.g we can assume that
      $c_m = x_1 \lor \neg x_2 \lor x_3$. This implies that
      $\{v_1,\bar{v}_2,v_3\}\cap Y = \varnothing$.
      From the construction of $\prec_\varphi$ we know that there are
      no elements dominating over $d_m$ except for
      $v_1,\bar{v}_2,v_3$. And since obviously $d_m \in X\setminus Y$,
      we receive $X \not \ll Y$ which is a contradiction.
  \end{enumerate}
\end{proof}

\noindent
Using the notion of \g-repairs also leads to a significant increase of
computational complexity when computing \g-preferred consistent query
answers. 
\begin{theorem}
  \label{thm:decl_answers_intractable}
  There exists a set of four functional dependencies $F$ and a
  quantifier-free ground query $\varphi$ (consisting of one atom only) 
  such that the set
  \[
   D^\g_{\varphi,F}=\{ (r,\mathord\prec) 
   | r \models_{F,\mathord\prec}^\g \varphi\}
  \]
  is $\Pi^p_2$-complete.
\end{theorem}
\begin{proof}
The membership of $D^g_{F,\varphi}$ in $\Pi_2^p$ follows from the
definition of \g-preferred consistent query answer: 
query is not \g-consistently true if it is false in some \g-repair,
and checking if a given set is a \g-repair is in co-NP. We show 
$\Pi_2^p$-hardness below.

Consider a quantified boolean formula $\psi$ of the form 
\begin{equation}
\psi = \forall x_1,\ldots{},x_n . \exists y_1,\ldots{},y_m . \phi,
\end{equation}
where $\phi$ is quantifier-free and is in 3CNF, i.e 
$\phi$ equals to $c_1 \land \ldots{} \land c_s$, and $c_k$ are clauses
of three literals $l_{k,1} \lor l_{k,2} \lor l_{k,3}$. We will
construct a database instance $r_\psi$ (over the schema
$R(A_1,B_1,\ldots{})$) and a priority relation $\prec_\psi$ 
such that true is a \g-preferred consistent answer to a query $R(Y)$
if and only if $\psi$ is true (the value of $Y$ can be found in Table
\ref{tab:tab5}). The set of integrity constraints is 
$C=\{A_1\rightarrow B_1,\ldots,A_4\rightarrow B_4\}$.   

We define two auxiliary functions $var$ and $sgn$ on literals in the
following fashion:
\begin{align*}
   var(x_i) = var(\neg x_i) & = i, & 
   sgn(x_i) = sgn(y_j) & = 1, \\
   var(y_j) = var(\neg y_j) & = n + j, &
   sgn(\neg x_i) = sgn (\neg y_j) & = -1.
\end{align*}

Now, we describe the tuples contained in $r_\psi$. 
\[
r_\psi=\{p_1,\bar{p}_1,\ldots{},p_n,\bar{p}_n,
q_1,\bar{q}_1,\ldots{},q_m,\bar{q}_m,
d_1\ldots{},d_s
\}.
\]
The exact values of tuples can be found in Table \ref{tab:tab5}.
\begin{table}[htb]\small
\begin{center}
  \[
  \begin{array}{|c|c|c|c|c|c|c|c|c|}
    \hline
               & A_1 & B_1 & A_2 & B_2 & A_3 & B_3 & A_4 & B_4 \\
    \hline
    \hline
     q_j       &  1  &  1  & n+j & -1  & n+j & -1  & n+j &  -1  \\
    \hline
     \bar{q}_j &  1  &  1  & n+j &  1  & n+j &  1  & n+j &  1   \\
    \hline
     Y         &  1  &  1  &  0  &  0  &  0  &  0  &  0  &  0   \\
    \hline
     X         &  1  &  2  &  0  &  0  &  0  &  0  &  0  &  0  \\
    \hline
     p_i       &  1  &  2  &  i  &  1  &  i  &  1  &  i  &  1   \\
    \hline
     \bar{p}_i &  1  &  2  &  i  &  -1 &  i  &  -1 &  i  &  -1  \\
    \hline
     d_k       &  1  &  2  & var(l_{k,1}) & sgn(l_{k,1}) & 
               var(l_{k,2}) & sgn(l_{k,2}) & var(l_{k,3}) &  sgn(l_{k,3})  \\
    \hline
    \end{array}
   \]
\end{center}
\caption{\label{tab:tab5} Values of tuples in $r_\psi$}
\end{table}
The priority relation $\prec_\psi$ is the unique minimal priority
relation that satisfies the following conditions:
\begin{align*}
  d_k & \prec_\psi p_i, & & 
  \text{if $c_k$ uses a positive literal $x_i$},\\
  d_k & \prec_\psi \bar{p}_i, & & 
  \text{if $c_k$ uses a negative literal $\neg x_i$},\\
  d_k & \prec_\psi q_j, & & 
  \text{if $c_k$ uses a positive literal $y_j$},\\
  d_k & \prec_\psi \bar{q}_j, & & 
  \text{if $c_k$ uses a negative literal $\neg y_j$},\\
  p_i & \prec_\psi Y,   & & \text{for all $i\in\{1,\ldots{},n\}$},\\
  \bar{p}_i & \prec_\psi Y,   & & \text{for all $i\in\{1,\ldots{},n\}$},\\
  X   & \prec_\psi Y. 
\end{align*}

In Figure \ref{fig:fig5} we can find a conflict graph of
an instance obtained from the reduction of a formula 
\[
\forall x_1, x_2, x_3. \exists y_1, y_2. 
(\neg x_1 \lor y_1 \lor x_2) \land 
(\neg x_2 \lor \neg y_2 \lor \neg x_3).
\]
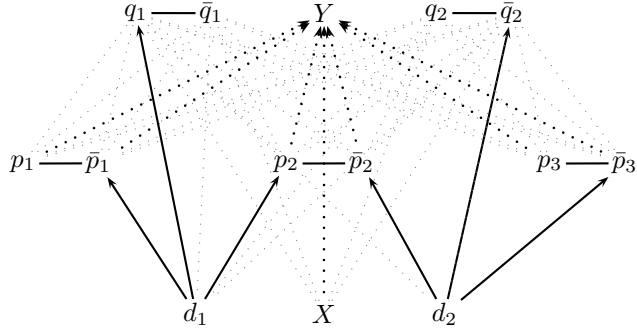
\begin{figure}[hbt]
\begin{center}
\begin{pspicture}(10,-4)(0,2)
  \rput(2.5,1){\rnode{q1}{$q_1$}}
  \rput(3.5,1){\rnode{_q1}{$\bar{q}_1$}}
  \rput(6.5,1){\rnode{q2}{$q_2$}}
  \rput(7.5,1){\rnode{_q2}{$\bar{q}_2$}}

  \ncline[nodesep=1.4pt]{-}{q1}{_q1}
  \ncline[nodesep=1.4pt]{-}{q2}{_q2}

  \rput(5,1){\rnode{Y}{$Y$}}
  \rput(5,-3){\rnode{X}{$X$}}

  \rput(1,-1){\rnode{p1}{$p_1$}}
  \rput(2,-1){\rnode{_p1}{$\bar{p}_1$}}
  \rput(4.5,-1){\rnode{p2}{$p_2$}}
  \rput(5.5,-1){\rnode{_p2}{$\bar{p}_2$}}
  \rput(8,-1){\rnode{p3}{$p_3$}}
  \rput(9,-1){\rnode{_p3}{$\bar{p}_3$}}

  \ncline[nodesep=1.4pt]{-}{p1}{_p1}
  \ncline[nodesep=1.4pt]{-}{p2}{_p2}
  \ncline[nodesep=1.4pt]{-}{p3}{_p3}

  \rput(3.3,-3){\rnode{d1}{$d_1$}}
  \rput(6.6,-3){\rnode{d2}{$d_2$}}

  \ncline[nodesep=1.4pt]{<-}{_p1}{d1}
  \ncline[nodesep=1.4pt]{<-}{q1}{d1}
  \ncline[nodesep=1.4pt]{<-}{p2}{d1}

  \ncline[nodesep=1.4pt]{<-}{_p2}{d2}
  \ncline[nodesep=1.4pt]{<-}{_q2}{d2}
  \ncline[nodesep=1.4pt]{<-}{_p3}{d2}





  \psset{linewidth=0.5pt,linecolor=gray,linestyle=dotted}

  \ncline[nodesep=1.4pt]{-}{q1}{p1}
  \ncline[nodesep=1.4pt]{-}{q1}{p2}
  \ncline[nodesep=1.4pt]{-}{q1}{p3}
  \ncline[nodesep=1.4pt]{-}{q1}{_p1}
  \ncline[nodesep=1.4pt]{-}{q1}{_p2}
  \ncline[nodesep=1.4pt]{-}{q1}{_p3}

  \ncline[nodesep=1.4pt]{-}{q2}{p1}
  \ncline[nodesep=1.4pt]{-}{q2}{p2}
  \ncline[nodesep=1.4pt]{-}{q2}{p3}
  \ncline[nodesep=1.4pt]{-}{q2}{_p1}
  \ncline[nodesep=1.4pt]{-}{q2}{_p2}
  \ncline[nodesep=1.4pt]{-}{q2}{_p3}

  \ncline[nodesep=1.4pt]{-}{_q1}{p1}
  \ncline[nodesep=1.4pt]{-}{_q1}{p2}
  \ncline[nodesep=1.4pt]{-}{_q1}{p3}
  \ncline[nodesep=1.4pt]{-}{_q1}{_p1}
  \ncline[nodesep=1.4pt]{-}{_q1}{_p2}
  \ncline[nodesep=1.4pt]{-}{_q1}{_p3}

  \ncline[nodesep=1.4pt]{-}{_q2}{p1}
  \ncline[nodesep=1.4pt]{-}{_q2}{p2}
  \ncline[nodesep=1.4pt]{-}{_q2}{p3}
  \ncline[nodesep=1.4pt]{-}{_q2}{_p1}
  \ncline[nodesep=1.4pt]{-}{_q2}{_p2}
  \ncline[nodesep=1.4pt]{-}{_q2}{_p3}

  \ncline[nodesep=1.4pt]{-}{_q1}{d1}
  \ncline[nodesep=1.4pt]{-}{q1}{d2}
  \ncline[nodesep=1.4pt]{-}{_q1}{2}

  \ncline[nodesep=1.4pt]{-}{q2}{d1}
  \ncline[nodesep=1.4pt]{-}{_q2}{d1}
  \ncline[nodesep=1.4pt]{-}{_q2}{d2}

  \ncline[nodesep=1.4pt]{-}{X}{q1}
  \ncline[nodesep=1.4pt]{-}{X}{_q1}
  \ncline[nodesep=1.4pt]{-}{X}{q2}
  \ncline[nodesep=1.4pt]{-}{X}{_q2}

  \psset{linewidth=1.0pt,linecolor=black,linestyle=dotted}
  \ncline[nodesep=1.4pt]{<-}{Y}{X}  
  \ncline[nodesep=1.4pt]{<-}{Y}{p1}
  \ncline[nodesep=1.4pt]{<-}{Y}{p2}
  \ncline[nodesep=1.4pt]{<-}{Y}{p3}
  \ncline[nodesep=1.4pt]{<-}{Y}{_p1}
  \ncline[nodesep=1.4pt]{<-}{Y}{_p2}
  \ncline[nodesep=1.4pt]{<-}{Y}{_p3}

\end{pspicture}
\end{center}
\caption{
\label{fig:fig5} 
     Conflict graph for
     $\forall x_1, x_2, x_3. \exists y_1, y_2. 
     (\neg x_1 \lor y_1 \lor x_2) \land 
     (\neg x_2 \lor \neg y_2 \lor \neg x_3)$ 
     and orientation $\prec_\psi$.
     The conflicts generated by $A_1 \rightarrow B_1$ are marked
     with dotted lines. 
}
\end{figure}

We partition the set of all repairs of $r_\psi$ into two (separate) 
classes: 
\begin{enumerate}
\item $\mathcal{Y}$-repairs: repairs that contain $Y$.
\item $\mathcal{X}$-repairs: repairs that don't contain $Y$. 
\end{enumerate}
We will use $\mathcal{X}$- and $\mathcal{Y}$-repairs to 'simulate' all 
possible valuations of variables $x_1,\ldots{},x_n$ and  
$y_1,\ldots{},y_m$ respectively. 

\subsubsection*{$\mathcal{Y}$-repairs}
Because of the functional dependency $A_1 \rightarrow B_1$ a repair is
$\mathcal{Y}$-repair if and only if it contains any of $q_j$ or
$\bar{q}_j$. Moreover for any $\mathcal{Y}$-repair $r'$ and for 
any $j$ either $q_j$ or $\bar{q}_j$ belongs to $r'$. Therefore there
is one-to-one correspondence between $\mathcal{Y}$-repairs and
valuations of $y_j$ variables. To easily move from the world of repairs
to the world of valuations and vice versa we define the following two
operators (for $r'$ being a $\mathcal{Y}$-repair and $V$ being a
valuation of variables in $\phi$):  
\[
V_\mathcal{Y}[r'](y_j) = 
\begin{cases}
true & q_j \in r'\\
false & \bar{q}_j \in r'
\end{cases} \qquad 
r_\mathcal{Y}[V] = \{q_j | V \models y_j\} \cup 
\{\bar{q}_j | V \models \neg y_j \} \cup \{Y\}.
\]

\subsubsection*{$\mathcal{X}$-repairs}
We will partition further the class of $\mathcal{X}$-repairs depending
on their 'conformance' with $\phi$. Because $\mathcal{X}$-repairs will
correspond only to valuations of $x_j$ we remove any usage of $y_j$
from $\psi$ in the following way: 
\begin{align*}
\tilde{y}_j = \neg \tilde{y}_j  &= false,\\
\tilde{x}_i & =  x_i, \\
\neg \tilde{x}_i &=  \neg x_i, \\
\tilde{c}_k & = 
\tilde{l}_{k,1} \lor \tilde{l}_{k,2} \lor \tilde{k}_{j,3},\\
\tilde{\phi} & = \tilde{c}_1 \land \ldots{} \land \tilde{c}_s.
\end{align*}
For a given valuation of $x_i$ construct the following set of tuples:
\[
r_\mathcal{X}[V] = \{p_i | V \models x_i\} \cup 
\{\bar{p}_i | V \models \neg x_i \}        \cup 
\{d_k | V \not \models \tilde{c}_k\}       \cup 
\{X\}.
\]
It's easy to verify that $r_\mathcal{X}[V]$ is a $\mathcal{X}$-repair. 
An $\mathcal{X}$-repair $r'$ is {\em strict} if and only if there
exists a valuation $V$ such that $r' = r_\mathcal{X}[V]$. Otherwise the
$\mathcal{X}$-repair is {\em non-strict}. 

It's clear that there is a one-to-one correspondence between strict
$\mathcal{X}$-repairs and valuations of $x_i$. Construction of
a valuation of $x_i$ from a strict $\mathcal{X}$-repair $r'$ is also
straightforward, for technical reasons we extend it to any
$\mathcal{X}$-repair:
\[
V_\mathcal{X}[r'](x_i) = 
\begin{cases}
true & p_i \in r'\\
false & \bar{p}_i \in r'\\
false & \text{otherwise}
\end{cases}
\]

Note that $\mathcal{X}$-repairs can be characterized in a alternative
way:
\begin{proposition}
A repair of $r_\psi$ is an $\mathcal{X}$-repair if and only if it
contains $X$.
\end{proposition}
In the main proof we use only strict $X$-repairs.  The following
observation will allow us to remove non-strict repairs from
consideration.  
\begin{claim} \label{cl:strict_are_maximal}
Strict $\mathcal{X}$-repairs are $\ll$-maximal
$\mathcal{X}$-repairs.
\end{claim}
\begin{proof} 
First we show how for any non-strict $\mathcal{X}$-repair $r'$
we construct a (strict) $\mathcal{X}$-repair $r''$ such that 
$r' \ll r''$. Take the valuation $V=V_\mathcal{X}[r']$ and let 
$r'' = r_\mathcal{X}[V]$. The repair $r''$ is strict and therefore
$r' \neq r''$. We show that $r'\ll r''$, i.e. 
\[
\forall t \in r' \setminus r''. 
\exists t' \in r'' \setminus r'. 
t \prec t'.
\]
There are three cases of values of $t$ to consider:
\begin{enumerate}
\item[$1^o$] $X\in r'\setminus r''$. Implies that $r''$ is not an
$\mathcal{X}$-repair, a contradiction.
\item[$2^o$] For some $i$ we have $p_i \in r'\setminus r''$ or
$\bar{p}_i \in r'\setminus r''$. W.l.o.g assume that 
$p_1\in r'\setminus r''$. This implies that $V(x_1) = true$. From
construction of $r_\mathcal{X}[V]$ this implies that $p_1 \in r''$, a
contradiction.
\item[$3^o$] For some $k$ we have $d_k\in r' \setminus r''$.
W.l.o.g. assume that $k=1$ and $c_1 = x_1 \lor y_1 \lor \neg
x_2$. Then $p_1 \not \in r'$ and $\bar{p}_2 \not \in r'$ (it's the
neighborhood of $d_1$). From the construction of $r''$ we have that 
\[
d_1\not \in r'' \iff 
V \not \models \tilde{c}_1 \iff
V \models x_1 \; \text{or} \; V \models \neg x_2 \iff 
p_1 \in r'' \; \text{or} \; \bar{p}_2 \in r''.
\]
And both $p_1$ and $\bar{p}_2$ dominate over $d_1$.
\end{enumerate}
Now, suppose that there exists a strict $\mathcal{X}$-repair $r'$ such
that there exists an $\mathcal{X}$-repair $r''$ preferred over $r'$. 
We show that $r'= r''$. Note that $r'$ and $r''$ must agree
on the tuples corresponding to the valuation of variables
$x_1,\ldots{},x_n$, i.e. 
\[
r' \cap \{p_1,\bar{p}_1,\ldots{},p_n,\bar{p}_n\} = 
r'' \cap \{p_1,\bar{p}_1,\ldots{},p_n,\bar{p}_n\}.
\]
Since $r'$ is strict, its content is determined by the
corresponding valuation of variables $x_1,\ldots{},x_n$. Therefore
$r'=r_\mathcal{X}[V_\mathcal{X}[r'']]$. We showed in the previous part
of the proof that $r'' \ll r'$. Since $\prec_\psi$ is acyclic this
implies that $r' = r''$. 
\end{proof}
\begin{claim}For any valuation $V$ of $x_i$ and $y_j$ we have 
$r_\mathcal{X}[V] \ll r_\mathcal{Y}[V]$ if and only if 
$V \models \phi$.
\end{claim}
\begin{proof} We prove implication in two directions:
\begin{itemize}
\item[\fbox{$\Leftarrow$}] By contradiction. Suppose $V\models\phi$  
and there exists a tuple $t$ of $r_\mathcal{X}[V]$ which is not
dominated by any tuple from $r_\mathcal{Y}[V]$. Obviously (from
dependency $A_1 \rightarrow B_1$) $t$ can be only one of
$d_k$. W.l.o.g. assume that $k=1$ and 
$c_1 = x_1 \lor y_1 \lor \neg x_2$. By construction of $r_\psi$ this
implies that $p_1 \not \in r_\mathcal{X}[V]$, 
$q_1 \not \in r_\mathcal{Y}[V]$, and 
$\bar{p}_2 \not \in r_\mathcal{X}[V]$. From the definition of 
$r_\mathcal{X}[V]$ and $r_\mathcal{Y}[V]$ we receive that 
$V(x_1) = false$, $V(x_2) = true$, and $V(y_1) = false$. This gives us 
$V \not \models c_i$ which is a contradiction. 
\item[\fbox{$\Rightarrow$}] By contradiction. 
Suppose $r_\mathcal{X}[V] \ll r_\mathcal{Y}[V]$ and there exists
conjunct $c_k$ such that $V \not \models c_k$. W.l.o.g. assume that 
$k=1$ and $c_1 = x_1 \lor y_1 \lor \neg x_2$. Then $V(x_1) = false$, 
$V(x_2) = true$, and $V(y_1) = false$. Consider $d_1$ and note that
it belongs to $r_\mathcal{X}[V]$ (by definition of
$r_\mathcal{X}$). From the construction of $\prec_\phi$ we know that
only $p_1$, $\bar{p}_2$, and $q_1$ dominate over $d_1$. 
$V_\mathcal{Y}[V]$ doesn't contain any of those and this gives us a
contradiction.
\end{itemize}
\end{proof}
\begin{proposition} QBF $\psi$ is true if and only if for any strict 
$\mathcal{X}$-repair $r'$ there exists a $\mathcal{Y}$-repair $r''$
such that $r' \ll r''$.
\end{proposition}
By Claim \ref{cl:strict_are_maximal} we have that only a
$\mathcal{Y}$-repair can be more preferred than a strict
$\mathcal{X}$-repair and for any non-strict $\mathcal{X}$-repair there
always exists a more preferred repair. 
\begin{corollary}
QBF $\psi$ is true if and only if for any $\mathcal{X}$-repair $r'$ 
there exists a different repair $r''$ such that $r' \ll r''$.
\end{corollary}
From the partition of repairs we know that $\mathcal{X}$-repairs can
be characterized with a formula $\neg R(Y)$. 
\begin{gather*}
\models \psi \iff 
\models \forall x_1,\ldots{},x_n . \exists y_1,\ldots{},y_m . \phi \iff\\
\forall r' \in \Rep_F(r_\psi). 
\left[r' \models \neg R(Y) \Rightarrow 
\exists r'' \in \Rep_F(r_\psi). r'\neq r'' \land 
r' \ll r''\right]. \iff\\
\forall r' \in \Rep_F(r_\psi).
[\neg \exists r'' \in \Rep_F(r_\psi). r'\neq r'' \land r' \ll r'']
\Rightarrow r' \models R(Y) \iff\\
\forall r' \in \GRep_F^{\prec_\psi}(r_\psi). 
r'\models R(Y) \iff\\
(r_\psi,\prec_\psi) \in D_{F,R(Y)}^g
\end{gather*}
\begin{corollary}
QBF $\psi$ is true if and only if true is \g-preferred consistent
answer to $R(Y)$ in $r_\psi$ w.r.t. $F$ and $\prec_\psi$.
\end{corollary}
If we use as characterization of $\mathcal{X}$-repairs the formula 
$R(X)$ then we can reduce QBF to answering to a query with one negated 
atom.
\begin{corollary}
QBF $\psi$ is true if and only if true is \g-preferred consistent
answer to $\neg R(X)$ in $r_\psi$ w.r.t. $F$ and $\prec_\psi$.
\end{corollary}
\end{proof}

\subsection{Database cleaning}
The postulate \ref{eq:p5} allows us to think of a total acyclic
priority as a cleaning program --- an exact specification of how to
resolve all conflicts. To run this program we simply use Algorithm
\ref{alg:alg1} and obtain a unique \l-preferred repair. Thanks to
Theorem \ref{thm:proc_equals_decl}, this is also the unique
\g-repair.
\begin{proposition} 
Given a a total acyclic priority $\prec$, the uni\-que \l-repair
(which if also the unique \g-repair) can be computed in time
polynomial in the size of the database. 
\end{proposition}

\section{Related work}
\label{sec:related}
We limit our discussion to work on using priorities to maintain
consistency and facilitate resolution of conflicts. 

The first to notice the importance of priorities in information
systems is \cite{FaUlVa83}. The authors study there the problem of
updates of databases containing propositional sentences. The priority
is expressed by storing a natural number with each clause (the
integrity constraints should be tagged with the highest priority
$0$). If an update (inserting or deleting a sentence) leads to
inconsistency, among all consistent and realizing the update databases
the minimally {\em different} are selected. A database $E$ is less
different than a database $F$ w.r.t. $D$ if either for some 
$i\in \{0,1,\ldots{},n\}$
\[
\left\{
\begin{array}{l}
D^{i-1} \setminus E^{i-1} = D^{i-1} \setminus F^{i-1},\\
D^i \setminus E^i \subset D^i \setminus F^i,
\end{array}
\right.
\qquad \text{or} \qquad
\left\{
\begin{array}{l}
D^n \setminus E^n = D^n \setminus F^n,\\
E  \setminus D \subset F \setminus D,
\end{array}
\right.
\]
where $n$ is the lowest priority in $D$ and $D^k$ consists of all
sentences from $D$ with priority less or equal to $k$. Although this
framework does not define a notion of a conflict, we note that more
than two facts can create a conflict w.r.t some constraint.
For sake of the comparison, assume that the conflicts are generated
only by pairs of facts (together with one of the constraints). Then,
the selected minimally different consistent databases are equivalent to
\g-repairs (and because the considered class of priorities has only
acyclic extensions it is equivalent to \l-repairs). We note, however, 
that the chosen representation of priorities imposes a significant
restriction on the class of considered priorities. In particular it
assumes transitivity of the priority on conflicting facts i.e. if
facts $a$, $b$, and $c$ are pair-wise conflicting and $a$ has a higher
priority  than $b$ and $b$ has a higher priority than $c$, then the
priority of $a$ is higher than $c$. This assumption cannot be always
fulfilled in the context of inconsistent databases. For example the
conflicts between $a$ and $b$, and between $b$ and $c$ may be caused
by violation of one integrity constraints while the conflict between
$a$ and $c$ is introduced by a different constraint. While the user
may supply us with a rule assigning priorities to conflicts created
by the first integrity constraint, the user may not wish to put any
priorities on any conflicts created by the other constraint.

A similar representation of priorities used to resolve inconsistency
in first-order theories is studied in \cite{Br89}, where the
inconsistent set of clauses is stratified (again the lowest strata has
the highest priority). Then preferred maximal consistent subtheories are
constructed in a manner analogous to \l-repairs. Furthermore, this
approach is generalized to priorities being a partial orders, by
considering all extensions to weak orders. Again, however, this
approach assumes transitivity of priority on conflicts, which as we 
explained previously may be considered a significant restriction. 

In \cite{Ry96} priorities are studied to facilitate the process of
{\em belief revision}. A belief state is represented as an ordered
list of propositional formulae and the revision operation simply adds
the given sentence at the end of the given belief state. This
representation of belief state allows to keep track of revision
history, which is later used to impose a preference order on the
possible interpretations of the belief state. Only maximally preferred
interpretations are used when defining the entailment relation.

In the context of logic programs, priorities among rules can be used to
handle inconsistent logic programs (where rules imply contradictory
facts). More preferred rules are satisfied, possibly at the cost of
violating less important ones. In a manner analogous to $\ll$,
\cite{NiVe02} lifts a total order on rules to a preference on
(extended) answers sets. When computing answers only maximally
preferred answers sets are considered.

\cite{SaIn00} investigate disjunctive logic programs with priorities
on facts. The authors use a transitive and reflexive closure (denoted
here $\preceq$) of a user supplied set of priorities on facts. The
preference on answer sets $\sqsubseteq$ is defined as follows: 
\begin{itemize}
\item $X \sqsubseteq X$ for every answer set $X$
\item $X \sqsubseteq Y$  if
\[
\exists y \in Y \setminus X. \Big[ \exists x \in X \setminus Y. x \preceq y 
  \land \neg \exists x' \in X \setminus Y. y \prec x' \Big],
\]
where $x\prec y$ stands for $x\preceq y \land y \not \preceq x$.
\item if $X \sqsubseteq Y$ and $Y \sqsubseteq Z$, 
then $X \sqsubseteq Z$.
\end{itemize}
The answer to a program in the extended framework consists of all
maximally preferred answer sets. The main shortcoming of using this
framework is it's computational infeasibility (which is specific to
decision problems involving general disjunctive programs): computing
answers to ground queries to disjunctive prioritized logic programs
under cautious (brave) semantics is $\Pi^p_3$-complete
(resp. $\Sigma_3^p$-complete).

A simpler approach to the problem of inconsistent logic programs is
presented in \cite{Gr97}. There conflicting facts are removed from
the model unless the priority specifies how to resolve the
conflict. Because only programs without disjunction are considered, 
this approach always returns exactly one model of the input
program. Constructing preferred repairs in a corresponding fashion
(by removing all conflicts unless the priority indicates a resolution)
would similarly return exactly one database instance
(fulfillment of \ref{eq:p1} and \ref{eq:p5}). However, if the priority 
does not specify how to resolve every conflict, the returned instance
is not a maximal set of tuples and therefore it is not a repair. 
Such an approach leads to a loss of (disjunctive) information and
violates postulates \ref{eq:p3} and \ref{eq:p4}.

\cite{FlGrZu04} proposes a framework of 
{\em conditioned active integrity constraints}, which allows the user
to specify the way some of the conflicts can be resolved. This notion
syntactically extends the notion  of embedded dependency 
$\forall X . [\phi \supset \exists Y  . \psi]$, 
where $X$ and $Y$  are sets of variables, $\phi$ and $\psi$ are two
conjunctions of literals, and each of existential variables $Y$ is
used only once. A conditioned active integrity constraint is obtained
by adding a disjunctive list of update atoms ($+C_1,\ldots{},+C_k$
for adding, and $-D_{k+1}, \ldots{}, -D_{n}$ for deletion) together
with conditions $\theta_1,\ldots{},\theta_n$ specifying when a
corresponding update atom can be used. Such an extended constraint is
denoted as  
\[
\forall X . [ (\phi \supset \exists Y . \psi) \supset 
 \theta_1 : +C_1 \lor \ldots{} \lor \theta_k : +C_k \lor
 \theta_{k+1} : -D_{k+1} \lor \ldots{} \lor \theta_n : -D_n]
\]
A constraint (or rather its grounded version) is said to be 
{\em  applied} to by a repair if the original integrity constraint
($\phi \subset \exists Y. \psi$) is satisfied in the database and the
repair is obtained by  performing updates satisfying the conditional
update atom lists (one of the atoms $C_1,\ldots{}.C_k$ has been added
and the corresponding condition $\theta_1,\ldots{},\theta_k$ is
satisfied, or one of the atoms $C_{k+1},\ldots{},C_n$ has been
removed and the corresponding condition
$\theta_{k+1},\ldots{},\theta_n$ is satisfied). On all repairs, which
are obtained in the standard way by taking as integrity constraints
only the heads of the conditioned action integrity constraints, we
define relation of preference: a repair $r_1$ is preferred over $r_2$
if every (ground) constraint applied in $r_1$ is also applied in
$r_2$. We note here that when restricted to functional dependencies
the set of preferred repairs is a superset of \l-repairs.  Inclusion
in the other direction doesn't always hold, which is illustrated on
the following example. 
\begin{example}
Consider a database $R(A_1,B_1,A_2,B_2)$ consisting of three tuples 
$r=\{t_1=(1,1,0,0),t_2=(1,2,3,3), t_3=(0,0,3,4)\}$
and suppose we work in the presence of two functional dependencies
$A_1\rightarrow B_1$ and $A_2 \rightarrow B_2$. Suppose also, that the
user specifies that if two tuples are conflicting w.r.t. the FD
$A_1\rightarrow B_1$, then the tuple with higher value of the field
$B_1$ should be preferred when repairing the database. A similar wish
is expressed for conflicts generated by the second functional
dependency. This can be expressed using the following two conditioned
active integrity constraints 
\begin{gather*}
\begin{split}
\forall x,y_1,y_2,z_1,z_2,s_1,s_2 . [
(R(x,y_1,z_1,s_1) \land R(x,y_2,z_2,s_2) \supset y_1 \neq y_2) \supset
\\
y_1>y_2 : -R(x,y_2,z_2,s_2)],
\end{split}
\\
\begin{split}
\forall x_1,x_2,y_1,y_2,z,s_1,s_2 . [
(R(x_1,y_1,z,s_1) \land R(x_2,y_2,z,s_2) \supset s_1 \neq s_2) 
\supset \\
s_1>s_2 : -R(x_2,y_2,z,s_2)].
\end{split}
\end{gather*}
After grounding we remove constraints with their head equal to false and we
obtain the following set 
\begin{align}
  R(1,1,0,0) \land R(1,2,3,3) \supset  1 > 2 : -R(1,2,3,3), \tag{I1} \label{eq:ic1}\\
  R(1,2,3,3) \land R(1,1,0,0) \supset  2 > 1 : -R(1,1,0,0), \tag{I2} \label{eq:ic2}\\
  R(1,2,3,3) \land R(0,0,3,4) \supset  3 > 4 : -R(0,0,3,4), \tag{I3} \label{eq:ic3}\\
  R(0,0,3,4) \land R(1,2,3,3) \supset  4 > 3 : -R(1,2,3,3). \tag{I4} \label{eq:ic4}
\end{align}
The corresponding priority relation is
${}\prec{}=\{(t_1,t_2),(t_2,t_3)\}$. Note that in the context of the
database $r$, the user has provided information sufficient to solve
all the conflicts, i.e. among the repairs 
$Rep_F(r) =\{r_1=\{t_1,t_3\},r_2 = \{t_2\}\}$ the repair $r_1$ is the
unique repair selected by $LRep_C^\prec$. At the same time only
\eqref{eq:ic2} is applied to $r_1$ and only \eqref{eq:ic4} is 
applied to $r_2$, what makes both repairs incomparable in terms
of the framework of \cite{FlGrZu04}.
\end{example}
This example also shows that the discussed framework violates the
postulate \ref{eq:p4}. Note also that removing preference information
on how to resolve the conflict between $t_2$ and $t_3$ will yield only
one repair $r_1$. This shows that this framework violates the
postulate \ref{eq:p5}. At the same time this framework fulfills the
property of conservativeness (the preferred repairs are a subset of
standard repairs) and non-emptiness (there is always at least one
preferred repair). 
\cite{FlGrZu04} also describes how to translate conditioned active
integrity constraints into a prioritized logic program \cite{SaIn00},
whose preferred models correspond to maximally preferred repairs. Note
that the framework of prioritized logic programming is computationally
more powerful (answering answers under the brave semantics is
$\Sigma_3^p$-complete) than required by the problem of finding if an
atom is present in any repair ($\Sigma_2^p$-complete). It is yet to be
seen if less powerful programming environment (like general disjunctive
logic programs) can be used to compute preferred answers. 

\cite{MoAnAc04} uses ranking functions on tuples to resolve conflicts
by taking only the tuple with highest rank and removing
others. This approach constructs a unique repair under the assumption
that no two different tuples are of equal rank (postulates
\ref{eq:p1} and \ref{eq:p5}). If this assumption is not satisfied and
the tuples contain numeric values, a new value, called the fusion, can
be calculated from the conflicting tuples (then, however, the
constructed instance is not a repair in the sense of Definition
\ref{def:repair}). 

A different approach based on ranking is studied in \cite{GrSiTrZu04}. 
The authors consider polynomial functions that are used to rank
repairs.  When computing preferred consistent query answers, only
repairs with the highest rank are considered. The postulates
\ref{eq:p1} and \ref{eq:p3} are trivially satisfied, but because this 
form of preference information does not have natural notions of
extensions and maximality, it is hard to discuss postulates \ref{eq:p4}
and \ref{eq:p5}. Also, the preference among repairs in this method is
not based on the way in which the conflicts are resolved. 

An approach where the user has a certain degree of control over the
way the conflicts are resolved is presented in \cite{GrLe04}. Using
repair constraints the user can restrict considered repairs to those
where tuples from one relation have been removed only if similar
tuples have been removed from some other relation. This approach is
monotonic, but not necessarily non-empty. The authors propose method
of weakening the repair constraints to restore non-emptiness, however
this comes at the price of losing monotonicity.

\section{Conclusions and future work}
\label{sec:conclusions}
In this paper we proposed a general framework of preferred repairs and
preferred consistent query answers by formulating a set of intuitive
postulates. We proposed two instantiations of the framework and
studied their semantic and computational properties. 
Table~\ref{tab:tab3} summarizes the computational complexity results;
its first row is taken from \cite{ChMa04}. 
\begin{table}[htb]{
\begin{center}
\begin{tabular}{|c|c|c|c|}
  \hline &  \multirow{3}{27pt}{Repair \\ Check}
               & \multicolumn{2}{c|}{Consistent Answers to}\\
  \cline{3-4}
         &   & $\{\forall,\exists\}$-free & conjunctive \\
         &   &  queries & queries \\
  \hline
  \hline
  \begin{tabular}{c}
    All repairs
  \end{tabular} & PTIME & PTIME & 
  co-NP-complete\\
  \hline
    \l-repairs & PTIME & \multicolumn{2}{c|}{co-NP-complete}\\
  \hline
    \g-repairs & co-NP-complete & \multicolumn{2}{c|}{$\Pi^p_2$-complete} \\
  \hline
\end{tabular}
\end{center}}
\caption{\label{tab:tab3} Summary of complexity results}
\end{table}

We envision several directions for further work. The postulates
\ref{eq:p1}--\ref{eq:p5} can be refined, so that only non-trivial
instantiations are captured. For example, the following instantiation
fulfills the postulates: we ignore any priority which is not total and
return all repairs in this case; when the priority is total we return
the unique \l-repair. This approach, however, is trivial and obviously
does not increase the computational complexity of any of considered 
problems. Also, the computational consequences of further refining the 
postulates should be examined. 

Along the lines of \cite{ABCHRS03}, the computational complexity
results could be further studied, by assuming a limit on the number of
functional dependencies or their conformance with BCNF.

The last is generalization of our framework to broader class of
constraints. Conflict graphs can be generalized to hypergraphs
\cite{ChMa04}, which allow to handle broader class of denial
constraints. Then, more than two tuples can be involved in a single
conflict and the current notion of priority does not have a clear
meaning.

\end{document}